\documentclass[11pt]{article}
\pdfoutput=1  
\usepackage{graphicx,color}
\usepackage{appendix}
\usepackage{latexsym,amsmath,amssymb,graphicx,booktabs}
\usepackage{epsfig,latexsym,cite}
\usepackage{hyperref}
\numberwithin{equation}{section}

\definecolor{MyBlue}{rgb}{0.15,0.15,0.70}

\hypersetup{
colorlinks=true,
citecolor=MyBlue,
linkcolor=MyBlue,
urlcolor=MyBlue
}

\input epsf
\usepackage{amssymb}
\usepackage{amsmath}
\usepackage{amsfonts}
\usepackage{upgreek}
\usepackage{latexsym}

\newcommand{\iBox}{\Box^{-1}}

\renewcommand\({\left(}
\renewcommand\){\right)}
\renewcommand\[{\left[}
\renewcommand\]{\right]}

\newcommand\n{{\mbox {\boldmath $\nabla$}}}
\newcommand{\ra}{\rightarrow}

\def\lsim{\raise 0.4ex\hbox{$<$}\kern -0.8em\lower 0.62
ex\hbox{$\sim$}}

\def\gsim{\raise 0.4ex\hbox{$>$}\kern -0.7em\lower 0.62
ex\hbox{$\sim$}}

\def\lbar{{\hbox{$\lambda$}\kern -0.7em\raise 0.6ex
\hbox{$-$}}}

\newcommand\eq[1]{eq.~(\ref{#1})}
\newcommand\eqs[2]{eqs.~(\ref{#1}) and (\ref{#2})}
\newcommand\Eq[1]{Equation~(\ref{#1})}

\newcommand\eqst[2]{eqs.~(\ref{#1})--(\ref{#2})}

\newcommand\pa{\partial}
\newcommand\p{\partial}

\newcommand\ee{\end{equation}}
\newcommand\be{\begin{equation}}
\def\bea{\begin{array}}
\def\eea{\end{array}}\def\ea{\end{array}}
\newcommand\ees{\end{eqnarray}}
\newcommand\bees{\begin{eqnarray}}
\def\nn{\nonumber}

\def\a{\alpha}
\def\b{\beta}

\def\g{\gamma}

\def\d{\delta}

\def\dslash{\hspace{-1mm}\not{\hbox{\kern-2pt $\partial$}}}
\def\Dslash{\not{\hbox{\kern-4pt $D$}}}
\def\pslash{\not{\hbox{\kern-2.1pt $p$}}}
\def\kslash{\not{\hbox{\kern-2.3pt $k$}}}
\def\qslash{\not{\hbox{\kern-2.3pt $q$}}}

\newcommand{\vk}{{\bf k}}
\newcommand{\vx}{{\bf x}}

\def\p1{{\bf p}_1}
\def\p2{{\bf p}_2}
\def\k1{{\bf k}_1}
\def\k2{{\bf k}_2}

\newcommand{\emn}{\eta_{\mu\nu}}

\newcommand{\gmn}{g_{\mu\nu}}

\newcommand{\gMN}{g^{\mu\nu}}

\newcommand{\pam}{\pa_{\mu}}

\newcommand{\pan}{\pa_{\nu}}

\newcommand{\paR}{\pa^{\rho}}
\newcommand{\paS}{\pa^{\sigma}}

\newcommand{\Gmn}{G_{\mu\nu}}

\newcommand{\Tmn}{T_{\mu\nu}}
\newcommand{\Smn}{S_{\mu\nu}}

\newcommand{\dddM}{\kern 0.2em \raise 1.9ex\hbox{$...$}\kern -1.0em \hbox{$M$}}
\newcommand{\dddQ}{\kern 0.2em \raise 1.9ex\hbox{$...$}\kern -1.0em \hbox{$Q$}}
\newcommand{\dddI}{\kern 0.2em \raise 1.9ex\hbox{$...$}\kern -1.0em\hbox{$I$}}
\newcommand{\dddJ}{\kern 0.2em \raise 1.9ex\hbox{$...$}\kern-1.0em
\hbox{$J$}}
\newcommand{\dddcalJ}{\kern 0.2em \raise 1.9ex\hbox{$...$}\kern-1.0em
\hbox{${\cal J}$}}

\newcommand{\dddO}{\kern 0.2em \raise 1.9ex\hbox{$...$}\kern -1.0em
\hbox{${\cal O}$}}
\def\dddz{\raise 1.5ex\hbox{$...$}\kern -0.8em \hbox{$z$}}
\def\dddd{\raise 1.8ex\hbox{$...$}\kern -0.8em \hbox{$d$}}
\def\dddbd{\raise 1.8ex\hbox{$...$}\kern -0.8em \hbox{${\bf d}$}}
\def\ddbd{\raise 1.8ex\hbox{$..$}\kern -0.8em \hbox{${\bf d}$}}
\def\dddx{\raise 1.6ex\hbox{$...$}\kern -0.8em \hbox{$x$}}

\newcommand{\ode}{\Omega_{\rm DE}}
\newcommand{\oma}{\Omega_{M}}
\newcommand{\ora}{\Omega_{R}}

\newcommand{\ola}{\Omega_{\Lambda}}

\newcommand{\rde}{\rho_{\rm DE}}

\newcommand{\rlam}{\rho_{\Lambda}}

\begin{document}

\begin{titlepage}

\vspace*{2cm}

\centerline{\Large \bf Cosmological dynamics and dark energy }

\vspace{5mm}

\centerline{\Large \bf  from non-local infrared modifications of gravity}

\vskip 0.4cm
\vskip 0.7cm
\centerline{\large Stefano Foffa, Michele Maggiore and Ermis Mitsou}
\vskip 0.3cm
\centerline{\em D\'epartement de Physique Th\'eorique and Center for Astroparticle Physics,}  
\centerline{\em Universit\'e de Gen\`eve, 24 quai Ansermet, CH--1211 Gen\`eve 4, Switzerland}

\vskip 1.9cm

\begin{abstract}
We perform a detailed study of the cosmological dynamics 
of a recently proposed infrared modification of the Einstein equations,
based on the introduction  of a  non-local term constructed with  
$m^2\gmn\Box^{-1} R$, where $m$ is a mass parameter.  
The theory   generates automatically a dynamical dark energy  component, that can reproduce the observed value of the dark energy density without introducing a cosmological constant. Fixing $m$ so to reproduce the observed value  
$\Omega_{\rm DE}\simeq 0.68$, and writing
$w(a)=w_0+(1-a) w_a$,
the  model provides a neat prediction for the equation of state parameters of dark energy, $w_0\simeq -1.042$ and $w_a\simeq -0.020$. We show that, because of some freedom in the definition of $\iBox$, one can extend the construction so to define  a more general family of non-local models. However, in a first approximation
this turns out to be equivalent to adding an explicit cosmological constant term on top of the dynamical dark energy component.
This leads to an extended model  with two parameters, $\ola$ and $m$.
Even in this extension the EOS parameter $w_0$ is always on
the phantom side,  in the range $-1.33\, \lsim\, w_0\leq -1$, 
and  there is a prediction for the relation between $w_0$ and $w_a$.

\end{abstract}


\end{titlepage}

\newpage

\section{Introduction}

The study of modifications of General Relativity (GR)  at cosmological scales has gained much impetus in recent years, as one of the most promising directions for understanding the origin of the observed acceleration of the Universe. The interest for such infrared (IR) modifications was initially spurred by the DGP model \cite{Dvali:2000hr}, which indeed has  a self-accelerating solution~\cite{Deffayet:2000uy,Deffayet:2001pu}. The viability of this specific proposal was eventually ruled out by the existence of a ghost instability~\cite{Luty:2003vm,Nicolis:2004qq,Gorbunov:2005zk,Charmousis:2006pn,Izumi:2006ca}, but the search for consistent IR modifications of GR and the study of their cosmological consequences
has been developed in various  different directions. In particular,  recent years have seen significant developments toward the construction of a consistent  theory of massive gravity 
\cite{deRham:2010ik,deRham:2010kj,deRham:2010gu,deRham:2011rn,deRham:2011qq,Hassan:2011hr,Hassan:2011vm,Hassan:2011tf,Hassan:2011ea,Hassan:2012qv,Comelli:2012vz,Jaccard:2012ut,Comelli:2013txa,Guarato:2013gba} (see \cite{Hinterbichler:2011tt} for  a review),
and the study of its cosmological consequences \cite{deRham:2010tw,Koyama:2011xz,Koyama:2011yg,Nieuwenhuizen:2011sq,Chamseddine:2011bu,D'Amico:2011jj,DeFelice:2013bxa,Tasinato:2013rza}.
Another aspect of this intense activity is that various  independent lines of reasoning seems to point  toward the relevance of some form of non-locality for the dark energy problem. Non-local operators that modify GR in the far IR appear in the degravitation proposal \cite{ArkaniHamed:2002fu,Dvali:2007kt} (see also ~\cite{Dvali:2000xg,Dvali:2002pe,Dvali:2006su}). A non-local cosmological model based on a non-local action has been  proposed in 
\cite{Deser:2007jk}, and  has been further studied in a number of recent papers, see e.g.~\cite{Koivisto:2008xfa,Koivisto:2008dh,Capozziello:2008gu,Elizalde:2011su,Zhang:2011uv,Elizalde:2012ja,Park:2012cp,Bamba:2012ky,Deser:2013uya,Ferreira:2013tqn,Dodelson:2013sma}. Another interesting non-local model  has been studied in 
\cite{Barvinsky:2003kg,Barvinsky:2011hd,Barvinsky:2011rk}. Non-local gravity models have also been studied as UV modifications of GR, see e.g.
\cite{Hamber:2005dw,Khoury:2006fg,Biswas:2010zk,Modesto:2011kw,Briscese:2012ys}.

In 
\cite{Jaccard:2013gla} it has been proposed a non-local modification of Einstein equation
of the form
\be\label{eqJMM}
\Gmn -m^2\(\iBox_{\rm ret}\Gmn\)^{\rm T}=8\pi G\,\Tmn\, .
\ee
We use the  notation $\Box$ to denote the d'Alembertian operator $\gMN\n_{\mu}\n_{\nu}$ with respect to the metric $\gmn$, and $\iBox_{\rm ret}$ is its inverse computed using the  retarded Green's function, as required by causality.
The superscript T denotes the extraction of the transverse part of the tensor, which exploits the fact that, in a generic curved space-time, any symmetric tensor $\Smn$ can be decomposed as
\be\label{splitSmn}
S_{\mu\nu}=S_{\mu\nu}^{\rm T}+\frac{1}{2}(\n_{\mu}S_{\nu}+\n_{\nu}S_{\mu})\, , 
\ee
where
$\n^{\mu}S_{\mu\nu}^{\rm T}=0$ \cite{Deser:1967zzb,York:1974}. The extraction of the transverse part of a tensor is a non-local operation. For instance in flat space, where $\n_{\mu}\ra\pam$, it is easy to show that the inversion of \eq{splitSmn} is
\be
S_{\mu\nu}^{\rm T}=\Smn
-\frac{1}{\Box}(\pam\paR S_{\rho\nu}+\pan\paR S_{\rho\mu})
+\frac{1}{\Box^2}\pam\pan\paR\paS S_{\rho\sigma}\, .
\ee
Because of its non-local nature,  the transverse part of a tensor does not  appear in the  classical equations of motion of a local theory. In \eq{eqJMM}, however, we already have an explicit $\iBox$ operator, so we have already payed the price of non-locality, and the use of the transverse part of a tensor becomes natural. Again, because of causality, we use the retarded Green's function to define the non-local operators than enter in the extraction of the transverse part.

\Eq{eqJMM} can be seen as a refinement of the original degravitation idea proposed in
\cite{ArkaniHamed:2002fu,Dvali:2007kt}, which was based on an equation of the form
\be\label{degorig}
\(1-\frac{m^2}{\Box}\)\Gmn=8\pi G\,\Tmn\, 
\ee
(with $m^2$ a constant or, more generally,  a function $m^2(\Box)$; this generalization could also be applied to \eq{eqJMM}, using an operator $m^2(\Box_{\rm ret})$ inside the transverse-part operation). A shortcoming of
\eq{degorig} is that, since the covariant derivative does not commute with $\Box^{-1}$, the left-hand side of \eq{degorig} is not transverse, and hence $\n^{\mu}\Tmn\neq 0$. In contrast, the left-hand side of \eq{eqJMM} is transverse by construction, so the energy-momentum tensor is automatically conserved. Observe furthermore that the use of the retarded Green's function in \eq{eqJMM} ensures causality.\footnote{In contrast, the original degravitation proposal 
\cite{ArkaniHamed:2002fu} was presented as an acausal modification of gravity at cosmological distances.} However , the presence of a retarded propagator already at the level of the equations of motion (rather than, as usual, just in their solution), has  important consequences for the conceptual meaning of such equations. As we discuss in detail in~\cite{FMMghost}, it implies that such non-local equations should not be understood as the equation of motion of a non-local QFT, but rather as  effective classical equations derived from some classical or quantum averaging of a more fundamental  local theory.

\Eq{eqJMM} can be further generalized to 
\be\label{modela1a2}
\Gmn -m^2\[ b_1\(\iBox_{\rm ret}\Gmn\)^{\rm T}
+b_2\frac{d-1}{2d}\(\gmn\iBox_{\rm ret}  R\)^{\rm T}\]=8\pi G\,\Tmn\, ,
\ee
where $b_1,b_2$ are arbitrary coefficients, and for the moment we work for generality in  $d$ spatial dimensions. The factor $(d-1)/(2d)$, is a convenient normalization of the $b_2$ coefficient. In particular, 
in \cite{Maggiore:2013mea}  has been studied the model with $b_1=0, b_2=1$, and it has been found that it has particularly interesting cosmological properties. 

The purpose of the  present paper is to elaborate in more detail on the cosmological results presented in  \cite{Maggiore:2013mea}. We will also discuss in some detail the consequences of the fact that different  definitions of the $\iBox$ operator are possible. Indeed,
the most general solution of an equation such as $\Box f=j$ is 
\be\label{fhom}
f(x)=(\iBox j)(x) \equiv f_{\rm hom}(x)+\int d^{d+1}x'\, \sqrt{-g(x')}\, G(x;x') j(x')\, ,
\ee
where $f_{\rm hom}(x)$ is any solution of $\Box f_{\rm hom}=0$ and $G(x;x')$ is any  Green's function of the $\Box$ operator. To define our non-local model we must specify what definition of $\iBox$ we use, i.e. we must specify the Green's function and the corresponding solution of the homogeneous equation. We will always use the retarded Green's function. Still, in a Friedmann-Robertson-Walker (FRW) spacetime, there remains a freedom due to the fact that there is no obvious initial time where the convolution with the Green's function starts. If we consider a model that, in the early Universe, starts from a radiation dominated (RD) phase, we can for instance start to convolution deep in RD  (e.g., even at $t=0$, as done  in
\cite{Deser:2007jk}). However, if we consider a model whose evolution begins in an earlier inflationary phase, the convolution will rather start at the beginning of the inflationary phase. Once extrapolated into the RD phase, this different definition of $\iBox_{\rm ret}$ will 
generate a non-vanishing homogeneous solutions, that depends on the earlier history.
As we will see, in FRW  in a first approximation this freedom turns out to be equivalent to the freedom of introducing an explicit cosmological constant term.  We will also discuss how the introduction of auxiliary fields  allows us to put these non-local models in  a local form. In this ``localized" form  the parameters labeling different definitions of the $\iBox$ operator, and hence different non-local theories, are mapped onto the initial conditions for the auxiliary fields. We will examine in detail the subtleties involved in this mapping which, if not properly taken into account, can easily lead to the inclusion of solutions that, with respect to a given initial non-local model, are spurious.

The organization of the paper is as follows.  In sect.~\ref{sect:iBoxR} we define our basic model and in sect.~\ref{sect:cd} 
we will perform a detailed analysis of its cosmological consequences,  expanding on the results presented in  
\cite{Maggiore:2013mea}. 
In  sects.~\ref{sect:iBoxR} and \ref{sect:cd} 
we  focus on a ``minimal" model, in which the evolution is started during RD  and the $\iBox$ operator is defined so that the associated
homogeneous solution in the RD phase is set to zero.  In sect.~\ref{sect:attr} we discuss how the definition of the $\iBox$ operator can be extended and we show that, when one writes the model in terms of auxiliary fields, this extension is reflected into the initial conditions of the auxiliary fields. Apart from allowing us to identify a more general class of models, the discussion in this section is important also for understanding the issue of the stability of the solution within a given non-local model. As we will see, the local formulation puts together the space of solutions of all these different models. As a result, apparent instabilities of a solution in the local formulation do not correspond necessarily to actual instabilities in the original non-local model, since they correspond to moving from the solution of a given non-local model toward the solutions of a different non-local model.

In a related paper~\cite{FMMghost} we discuss  in greater generality the conceptual issues raised by these non-local equations, in particular in connection with apparent ghost-like degrees of freedom that seem to emerge from these models, and we  show that such  apparent ghosts are spurious and do not represent propagating  degrees of freedom of the theory.
Models of the form (\ref{modela1a2}) with $b_1\neq 0$ seem less viable because of cosmological instabilities, and we examine them in App.~\ref{sect:iGmn}.
Our notation and convention are as in \cite{Jaccard:2013gla}. In particular, we use the signature $\emn=(-,+,+,+)$.

\section{The ``minimal" model}\label{sect:iBoxR}

We now set   $b_1=0, b_2=1$ in \eq{modela1a2}, i.e. we study the model given by  
\be\label{modela1b2}
\Gmn -m^2\frac{d-1}{2d}\(\gmn \iBox_{\rm ret}  R\)^{\rm T}=8\pi G\,\Tmn\, .
\ee
First of all, we need to give a precise definition of the $\iBox$ operator, i.e. we must assign the Green's function and the corresponding homogeneous solution in \eq{fhom}.
We directly  specialize to a spatially flat  FRW metric in $d$ spatial dimensions, $ds^2=-dt^2+a^2(t)d\vx^2$. In this section we follow \cite{Deser:2007jk} and we define
\be\label{iBoxDW}
(\iBox_{\rm ret}  R)(t)=-\int_{t_*}^{t} dt'\, \frac{1}{a^d(t')}
\int_{t_*}^{t'}dt''\, a^d(t'') R(t'')\, ,
\ee
where $t_*$ is some initial value of time, that we take here in RD. 
As we discuss in \cite{FMMghost}, a non-local equation such as (\ref{modela1b2}), which involves the retarded inverse d'Alembertian, should be understood as an effective equation, obtained from some classical or quantum  averaging of an underlying fundamental theory. Then, $t_*$ can be interpreted as a value of time where such an effective description becomes appropriate, and \eq{iBoxDW} is only valid for $t>t_*$.
Observe that, since in RD the Ricci scalar $R$ vanishes, this definition is independent of the exact value of $t_*$, as long as it is deep in RD. With this definition,  also $\iBox_{\rm ret}  R$ vanishes during RD, and only becomes active in the subsequent matter dominated (MD) phase.
In  FRW, on a scalar $f(t)$,  we have $\Box f=-a^{-d}\pa_0(a^d\pa_0f)$, so one immediately verifies that \eq{iBoxDW} indeed provides a possible inversion of the $\iBox$ operator.
This inversion corresponds to a retarded Green's function, as we see from the fact that the integration is only over times $t''$ and $t'$ smaller than $t$. Equivalently, we can rewrite \eq{iBoxDW} as
\be
(\iBox_{\rm ret}  R)(t)=-\int_{t_*}^{\infty} dt' \theta (t-t')\frac{1}{a^d(t')}
\int_{t_*}^{\infty}dt'' \, \theta (t'-t'')a^d(t'') R(t'')\, ,
\ee
which can be rearranged in the form
\be\label{iBoxGret}
(\iBox_{\rm ret}  R)(t)=\int_{t_*}^{\infty} dt'\,  G_{\rm ret}(t;t') R(t')\, ,
\ee
where
\be\label{Gret}
G_{\rm ret}(t;t') =-\theta(t-t') a^d(t')\int_{t'}^{t} dt'' \, \frac{1}{a^d(t'')}\, .
\ee
In sect.~\ref{sect:attr} we will study a more general class of models, in which we add a general solution of the homogeneous equations to the definition
(\ref{iBoxDW}) and we will find, quite remarkably, that the above freedom basically amounts to the possibility of introducing in the theory a cosmological constant term.

A similar issue of definition of non-local operators arises when we  compute the transverse part 
in \eq{modela1b2}. To extract the transverse part
we proceed as in \cite{Jaccard:2013gla,Maggiore:2013mea}.
We introduce a scalar field $U$ from 
\be\label{Udef}
U\equiv -\iBox_{\rm ret} R \equiv 
\int_{t_*}^{t} dt'\, \frac{1}{a^d(t')}
\int_{t_*}^{t'}dt''\, a^d(t'') R(t'')\, . 
\ee
We then define $\Smn=-U\gmn$, and
we split $S_{\mu\nu}$  as in \eq{splitSmn}. 
To determine  $S_{\mu}$  we apply $\n^{\mu}$ to both sides of this equation,
obtaining
\be\label{panU}
\Box S_{\nu}+\n^{\mu}\n_{\nu}S_{\mu}=-2\pan U\, .
\ee
We must therefore invert the operator $(\d^{\mu}_{\nu}\Box+\n^{\mu}\n_{\nu})$.
In FRW this inversion simplifies considerably. Indeed, the three-vector $S^i$  vanishes because there is no preferred spatial direction, while from the $\nu=0$ component  of \eq{panU} we get a differential equation for  $S_0$,
\be\label{eqS0}
\ddot{S}_0+dH\dot{S}_0-dH^2S_0
=\dot{U}\, . 
\ee
In this case we must therefore invert the  operator
\be
{\cal D}=\pa_0^2+dH\pa_0-dH^2\, . 
\ee
Denoting by $D_{\rm ret}(t;t')$ the retarded Green's function of this operator, the definition analogous to (\ref{iBoxGret}) is 
\be\label{S0Dret}
S_0(t)=\int_{t_*}^{\infty} dt'\,  D_{\rm ret}(t;t') \dot{U}(t')\, ,
\ee
i.e. we set again to zero the solutions of the associated homogeneous equation ${\cal D}f=0$. We will refer to the non-local model that makes use of these definitions of
$\iBox_{\rm ret}$ and ${\cal D}^{-1}_{\rm ret}$ as the ``minimal model".

We can now write down the cosmological equations governing this model. Since the energy-momentum tensor in \eq{modela1b2} is conserved by construction,
the cosmological evolution is determined by the $(0,0)$ component of \eq{modela1b2}, i.e. the Friedmann equation. 
\be\label{eq2}
H^2-\frac{m^2}{d^2}(U-\dot{S}_0)=\frac{16\pi G}{d(d-1)}\rho\, .
\ee
More explicitly, inserting the definitions (\ref{iBoxGret}) and
(\ref{S0Dret}), our non-local model is defined by the integro-differential equation
\bees
&&H^2+\frac{m^2}{d^2}\[ \int_{t_*}^{\infty} dt'\,  G_{\rm ret}(t;t') R(t')
-\pa_t\int_{t_*}^{\infty} dt'\,  D_{\rm ret}(t;t') \pa_{t'}
\int_{t_*}^{\infty} dt''\,  G_{\rm ret}(t';t'') R(t'')\]\nn\\
&&=\frac{16\pi G}{d(d-1)}\rho\, .\label{originteg}
\ees

\section{Cosmological dynamics}\label{sect:cd}

\subsection{Local form of the evolution equations}

To evolve the equation numerically it can be convenient to transform the  integro-differential equation (\ref{modela1b2}) into a set of local equations. This  can be obtained using the auxiliary fields $U(t)$ and $S_0(t)$ defined
above. \Eq{Udef} can be written as $\Box U=-R$ so, together with
\eqs{eqS0}{eq2},   we have three differential
equations for the three functions $\{H(t),U(t),S_0(t)\}$.
The retarded prescriptions in \eqs{Udef}{S0Dret} are automatically taken into account by assigning initial conditions on $U(t)$ and $S_0(t)$ at an initial time $t_{*}$ and integrating the equations forward in time. 

To integrate, we must  then assign 
$U,\dot{U}, S_0$ and $\dot{S}_0$ at $t=t_*$. In turn, these initial conditions are uniquely specified by the definitions of the $\iBox_{\rm ret}$ and 
${\cal D}^{-1}_{\rm ret}$ operators given in \eqs{Udef}{S0Dret}, and in particular by the choice of the associated homogeneous solutions, which here we have set to zero. Thus,
from \eq{Udef} we have $U(t_*)=0$. Furthermore, \eq{Udef} gives
\be
\dot{U}(t)=\frac{1}{a^d(t)}
\int_{t_*}^{t}dt''\, a^d(t'') R(t'')\, ,
\ee
and therefore also $\dot{U}(t_*)=0$. Similarly, the retarded nature of 
$D_{\rm ret}(t;t')$ in \eq{S0Dret} implies that $S_0(t_*)=0$. Furthermore, writing
$D_{\rm ret}(t;t')=\theta(t-t') g(t;t')$, we have
\bees
\dot{S}_0(t)&=&\int_{t_*}^{\infty} dt'\,  \[\d(t-t')g(t;t')+\theta(t-t') \pa_t g(t;t')\]
\dot{U}(t')\nn\\
&=&g(t;t)\dot{U}(t)+\int_{t_*}^{t} dt'\, \pa_t g(t;t')\dot{U}(t')\, .
\ees
In $t=t_*$ this vanishes,  because $\dot{U}(t_*)=0$. In summary, the original integro-differential equation (\ref{originteg}) is equivalent to the coupled system of differential equations
\bees
H^2-\frac{m^2}{d^2}(U-\dot{S}_0)&=&\frac{16\pi G}{d(d-1)}\rho\,
\label{loc1} \\
\ddot{U}+dH\dot{U}&=&2d\dot{H}+d(d+1)H^2\, ,\label{loc2}\\
\ddot{S}_0+dH\dot{S}_0-dH^2S_0&=&\dot{U}\, ,\label{loc3}
\ees
(where we used the fact that, in FRW with generic $d$, 
$R=2d\dot{H}+d(d+1)H^2$),
together with the initial conditions
\be\label{initcond}
U(t_*)=\dot{U}(t_*)=S_0(t_*)=\dot{S}_0(t_*)=0\, .
\ee
It is important to stress that the initial conditions on the auxiliary fields $U$ and $S_0$ are fixed, once  we give the definition of the $\iBox$ and ${\cal D}^{-1}$ operators in the original non-local model. Taking these initial conditions as free parameters is incorrect. In other words, the space of solutions of the local system 
(\ref{loc1})--(\ref{loc3}), with generic initial conditions on $U$ and $S_0$, is much larger than the space of solutions of the original non-local equation. Different
choice of initial conditions on $U$ and $S_0$ correspond to different choices of the homogeneous solutions associated to \eqs{loc2}{loc3}, i.e. of the equations 
$\Box U=0$ and ${\cal D}S_0=0$, which corresponds to different choices of the homogeneous functions used to define the $\iBox_{\rm ret}$ and
${\cal D}^{-1}_{\rm ret}$  in the original non-local model. Any given definition of
$\iBox_{\rm ret}$ and ${\cal D}^{-1}_{\rm ret}$ fixes a corresponding solution of the homogeneous solutions associated to \eqs{loc2}{loc3}. If one forgets this simple but important point, one can easily fall into the mistake of believing that the solutions of 
$\Box U=0$ and ${\cal D}S_0=0$ represent scalar propagating degrees of freedom of the original non-local model.
The fact that these degrees of freedom are spurious, and are an artifact of the ``localization" procedure, has been recognized recently by various authors in similar non-local models
\cite{Koshelev:2008ie,Koivisto:2009jn,Barvinsky:2011rk,Deser:2013uya}.
The issue is even more important  in flat Minkowski space, where these spurious degrees of freedom include a ghost. This would lead to the erroneous conclusion that the quantum vacuum of these theories is unstable. In fact,  there is no propagating degree of freedom associated to the ghost. In the flat-space case the solutions of the associated homogeneous equation $\Box U=0$ are of course just plane wave. However, the coefficients $a_{\vk}$ and $a^*_{\vk}$ of these plane-wave solutions are not free parameters that, at the quantum level, can be promoted to annihilation and creation operators of a quantum field. Simply, they are fixed once the definition of the $\iBox$ operator is given (e.g. to $a_{\vk}=a^*_{\vk}=0$), and do not parametrize degrees of freedom of the original non-local theory (see also the more extended discussion in \cite{FMMghost}).

\subsection{Cosmological evolution}

Having clarified this important conceptual point, we can now use the local form of the equations to study the cosmological evolution.
We take   $\rho$ equal to the sum of the matter density $\rho_M$ and the radiation density $\rho_R$, and we henceforth restrict to $d=3$ spatial dimensions. We do not add by hand a cosmological constant term $\rho_{\Lambda}$, since  our aim is to investigate whether a viable dynamical dark energy (DE) component emerges automatically from the term proportional to the  mass $m$. It is also convenient to define $Y=U-\dot{S}_0$, since this is the quantity that appears in \eq{eq2}, and use
$\{H,U,Y\}$ as independent variables.
We also define 
\be\label{defrhode}
\rde(t)=\rho_0\g Y(x)\, , 
\ee
where $\rho_0=3H_0^2/(8\pi G)$, and
\be
\g\equiv  \frac{m^2}{9H_0^2}\, .
\ee
Then
\eq{eq2} becomes
\be\label {H2DE}
H^2(t)=\frac{8\pi G}{3}\[\rho_M(t)+\rho_R(t)+\rho_{\rm DE}(t)\]\, .
\ee
Thus,  the term proportional to $m^2$ plays the role of a dynamical dark energy.
In order  to deal with dimensionless quantities only 
we define  as usual $h(t)=H(t)/H_0$,
$\Omega_i (t)=\rho_i(t)/\rho_c(t)$ (where $\rho_c(t)=3H^2(t)/(8\pi G)$ and
$i$ labels radiation, matter and dark energy), and we use the notations $\Omega_M\equiv \Omega_M (t_0)$, $\Omega_R\equiv \Omega_R (t_0)$, $\Omega_{\rm DE}\equiv \Omega_{\rm DE} (t_0)$.   We find useful to parametrize the temporal evolution using the variable
$x\equiv \ln a(t)$ instead of $t$, and we denote $df/dx=f'$. 
Then, we  get \cite{Maggiore:2013mea}
\be\label{hLCDM}
h^2(x)=\Omega_M e^{-3x}+\Omega_R e^{-4x}+\g Y(x)\, ,
\ee
where the evolution of $Y(x)$ is obtained from the coupled system of equations
\bees
&&Y''+(3-\zeta)Y'-3(1+\zeta)Y=3U'-3(1+\zeta)U\, ,\label{sy1R}\\
&&U''+(3+\zeta)U'=6(2+\zeta)\label{sy3R}\, ,
\ees
and $\zeta$ is  given by 
\be\label{syzR}
\zeta(x)\equiv\frac{h'}{h}=-\, \,
\frac{3\Omega_M e^{-3x}+4\Omega_R e^{-4x}
-\g Y' }{2(\Omega_M e^{-3x}+\Omega_R e^{-4x}+\g Y)}\, .
\ee
The initial  conditions (\ref{initcond}), together with the definition $Y=U-\dot{S}_0$, imply that $Y(t_*)=0$. Furthermore, using \eq{loc3}, we see that \eq{initcond} also implies that $\ddot{S}_0(t_*)=0$, and therefore also $\dot{Y}(t_*)=0$. Thus, the initial conditions corresponding to the original integro-differential equation
 (\ref{originteg}) are
 \be\label{ini2}
 U(t_*)=U'(t_*)=Y(t_*)=Y'(t_*)
 =0\, .
\ee
Observe  that $\zeta(x)$ is related to the total equation of state (EOS) parameter $w(t)$, defined by $p(t)=w(t)\rho(t)$, where
$p=\sum_i p_i$, $\rho=\sum_i \rho_i$
(and, again, $i$ labels radiation, matter and dark energy). Combining energy-momentum conservation 
$\dot{\rho}+3(1+w)H\rho=0$ with the Friedmann equation
$H^2=(8\pi G/3)\rho$ we get in fact $\dot{H}/H^2=-(3/2)[1+w(t)]$ or,
using $x$ as time evolution variable and observing that  $\dot{H}/H^2=H'/H$,
\be\label{zetaEOS}
\zeta (x)=-\frac{3}{2}[1+w(x)]\, .
\ee
We finally define the dark energy equation-of-state (EOS) parameter $w_{\rm DE}(x)$ from
\be\label{defwDE}
\dot{\rho}_{\rm DE}+3(1+w_{\rm DE})H\rho_{\rm DE}=0\, .
\ee
Observing that  
$\dot{\rho}=H\rho'$ we get
\be\label{wg2}
w_{\rm DE}(x)=-1 -\frac{Y'(x)}{3Y(x)}\, .
\ee
The EOS parameter of this dark energy component is therefore close to $-1$ if
$|Y'/3Y|\ll 1$.

\subsection{Perturbative solutions and stability}\label{sect:pertstab}

The above equations are highly non-linear, because the function $Y(x)$ and its derivative appears also in $\zeta(x)$. 
As discussed in \cite{Maggiore:2013mea} it is useful to begin by studying a perturbative regime, where the contribution of $Y(x)$ to $\zeta(x)$   is  negligible. In particular we expect that this will be true in the early Universe (i.e.  at $x$ large and negative) so that we recover standard cosmology at early times. We therefore assume that, as $x\ra-\infty$, 
\be
\zeta(x)\simeq -\, \,
\frac{3\Omega_M e^{-3x}+4\Omega_R e^{-4x} }{2(\Omega_M e^{-3x}+\Omega_R e^{-4x})}\, ,
\ee
and we  check a posteriori the self-consistency of the procedure.
In this case, in each given era $\zeta(x)$ can be further approximated by a constant $\zeta_0$, with $\zeta_0=-2$ in RD and $\zeta_0=-3/2$ in MD, and \eq{sy3R} can be integrated analytically. 
The perturbative solution for $U$ is  given by \cite{Maggiore:2013mea} 
 \be \label{pertU}
U(x)=\frac{6(2+\zeta_0)}{3+\zeta_0}x+u_0
+u_1 e^{-(3+\zeta_0)x}\, ,
\ee
where the coefficients $u_0,u_1$ parametrize the general solution of the homogeneous equation $U''+(3+\zeta_0)U=0$. For later use, we study  here the perturbative solution with generic initial conditions, and we will later impose the initial conditions (\ref{ini2}) appropriate to our problem.
Plugging \eq{pertU} into \eq{sy1R} and solving for $Y(x)$ 
we get \cite{Maggiore:2013mea}
\bees
Y(x)&=&-\frac{2(2+\zeta_0)\zeta_0}{(3+\zeta_0)(1+\zeta_0)}
+\frac{6(2+\zeta_0)}{3+\zeta_0} x+u_0
-\frac{6(2+\zeta_0) u_1 }{2\zeta_0^2+3\zeta_0-3}e^{-(3+\zeta_0)x} 
\nn\\
&& +a_1 e^{\a_{+}x}+ a_2 e^{\a_{-}x}\, ,\label{pertY}
\ees
where 
\be\label{apm}
\a_{\pm}=\frac{1}{2}\[-3+\zeta_0\pm \sqrt{21+6\zeta_0+\zeta_0^2}\]\, .
\ee
Observe that in RD $\zeta_0=-2$ and the inhomogeneous solutions for $U$ and $Y$ vanish. This is a consequence of the fact that in RD the Ricci scalar vanishes, so  $\Box U=0$ and the only contributions to $U$ and to $(U\gmn)^T$ come from the solutions of the homogeneous equations. The  inhomogeneous solution  is self-consistent with our perturbative approach. Indeed, in a pure RD phase it just vanishes, and in a generic epoch, as $x\ra-\infty$, $Y(x)\propto x$ so its contribution to $\zeta(x)$  is  anyhow negligible compared to the term $\oma e^{-3x}$ and $\ora e^{-4x}$ in \eq{syzR}. 

Specializing now the case in which the evolution is started at a value $x=x_*$ deep in RD, we see that at the initial time the inhomogeneous solution vanishes and therefore\bees \label{pertU*}
U(x_*)&=&u_0+u_1 e^{-x_*}\, ,\\
Y(x_*)&=&u_0+a_1 e^{\a_{+}x_*}+ a_2 e^{\a_{-}x_*}\, .
\ees
Imposing the initial conditions (\ref{ini2}) in RD therefore amounts to setting
$u_0=u_1=a_1=a_2=0$, i.e. we set to zero the solution of the homogeneous equations in RD.

In sect.~\ref{sect:attr} we will study what happens if we rather start the evolution in an earlier phase, such as an earlier inflationary epoch.
Observe that, in a generic epoch, the homogeneous solutions for $U$ are  always stable (as long as $\zeta_0\geq -3$, i.e. $w\leq 1$, which is always the case). The homogeneous solution for $Y$ is stable as long as both $\a_{+}\leq 0$ and $\a_{-}\leq 0$. This gives the condition $\zeta_0\leq -1$, i.e.
\be\label{w013}
w_0\equiv -1-\frac{2}{3}\zeta_0\geq -\frac{1}{3}\, ,
\ee
which is satisfied in RD and MD.  In particular, in RD
$\a_{\pm}=(1/2)(-5\pm\sqrt{13})$ and in MD $\a_{\pm}=(-9\pm\sqrt{57})/4$.
However, the condition
$w_0<-1/3$ is the condition for having an accelerated expansion so, if we start the evolution in an inflationary era, and we allow for  generic values of the coefficients
$a_1, a_2$,  the perturbative solution is unstable. However, as discussed above (and as we will discuss again in detail in sect.~\ref{sect:attr}) the initial conditions are in one-to-one correspondence with the definition of the non-local operators in the original non-local model. Thus, if we start the evolution in an earlier inflationary era, and we define the  non-local operators $\iBox_{\rm ret}$ and ${\cal D}^{-1}_{\rm ret}$ so that their associated homogeneous solutions vanish, we must set $a_1=a_2=0$ in 
\eq{pertY} in the perturbative solution valid during the inflationary era. With this definition of the non-local model, the exponentially growing homogeneous solutions are simply not solutions of the original non-local integro-differential equation, and are  an artifact due to the fact that the space of solutions of the local form of the equations is larger than the space of solutions of the original non-local model.
In turn, setting to zero the homogeneous solutions during the inflationary era will generate non-zero homogeneous solutions during the subsequent RD era, whose effect will be studied in sect.~\ref{sect:attr}.

\subsection{Numerical solution of the full equations}\label{sect:num}

We now integrate \eqst{hLCDM}{syzR}  numerically.  Since the initial conditions on $U,Y$ are fixed by the definition of $\iBox_{\rm ret}$ and ${\cal D}^{-1}_{\rm ret}$
in the original non-local model, the only free parameter is $\gamma$, plus of course the values of $\oma$ and $\ora$ that enter through \eq{hLCDM}.  However, just as in $\Lambda$CDM the parameters $\ola,\oma$ and $\ora$ are related by the condition 
$\oma+\ora+\ola=1$, similarly here $\gamma,\oma$ and $\ora$ are related by the condition that, at $x=0$, $\oma+\ora+\gamma Y(0)=1$.
In other words, since by definition $h(x)=H(x)/H_0$, the only consistent solutions are those  that  satisfy $h(0)=1$. We set $\oma$ and $\ora$  to
the Planck best-fit  values   $\oma=0.3175$,
$\ora=4.15\times 10^{-5}h_0^{-2}$, $h_0=0.6711$ \cite{Ade:2013zuv} (and we set $\ola=0$). 
The appropriate value of $\gamma$ must then be determined by trials and errors, since $\ode=\gamma Y(0)$ and the evolution of $Y(x)$ depends on $\gamma$ itself through the dependence of $\zeta(x)$ on $\gamma$.\footnote{Alternatively we could start from the equations in the form (\ref{loc1})--(\ref{loc3}), fix an initial value $x_*$, say deep in RD, assign $\gamma$ as well as the values of $\rho_M(x_*)$ and $\rho_R(x_*)$, and let the system evolve forward in time (again, the initial conditions on $U$ and $S_0$ are uniquely fixed by the definition of the non-local operators in the original non-local model). 
The present value of time $t_0$ (or, equivalently,  the value $x_0$) is then identified by the condition that $H(x)$ reaches the observed value $H_0$. Each value of 
$\{ \gamma, \rho_M(x_*),\rho_R(x_*)\}$ produces a given matter and dark energy content at $x=x_0$.
The values of $\gamma$, $\rho_M(x_*)$ and $\rho_R(x_*)$ could then be chosen, by trial and errors, so to obtain the desired values of $\oma$ and $\ora$ today. 
However, passing to the dimensionless quantity $h(x)$ and fixing  directly $\oma$ and $\ora$ in  \eq{hLCDM} to the desired values is a much more effective way of proceeding, since then we must vary just a single parameter $\gamma$. The reason is that in the three-dimensional space spanned by $\gamma$, $\rho_M(x_*)$ and $\rho_R(x_*)$ there are degeneracies, due to the fact that two models with different values of
$\{ \gamma, \rho_M(x_*),\rho_R(x_*)\}$ can reach the same values of $H_0$ at different values of $x_0$.  Imposing that the present time is at $x_0=0$ removes this degeneracy.}
 We find that, having  set $\oma = 0.3175$, $h_0 = 0.6711$ and  $\ora = 4.15\times 10^{-5} h_0^{-2}$, the required value  is $\gamma =0.050255$ (where this number of digits is necessary so  that $|\oma+\ora+\ode-1|<10^{-4}$). This corresponds to
$m/H_0=3\gamma^{1/2} \simeq 0.674 H_0$.  

\begin{figure}[t]
\begin{center}
\begin{minipage}{1.\linewidth}
\centering
\includegraphics[width=0.45\columnwidth]{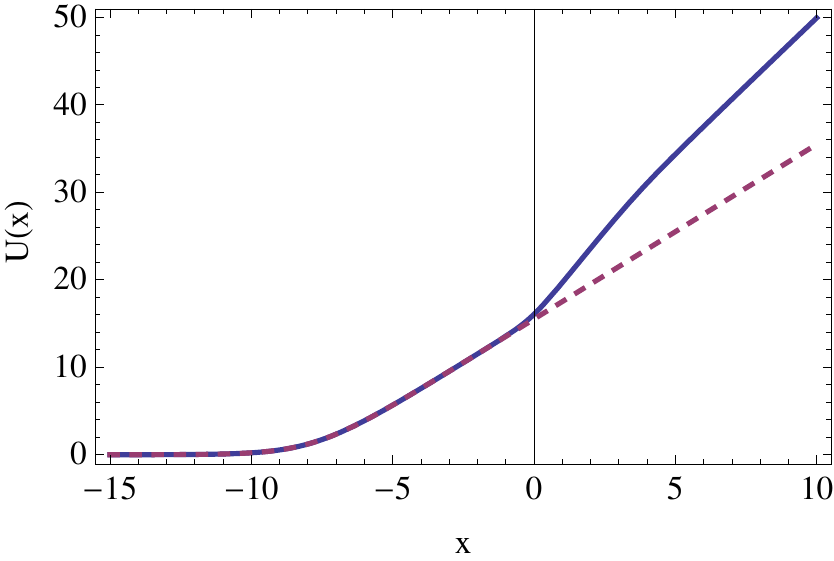}
\includegraphics[width=0.45\columnwidth]{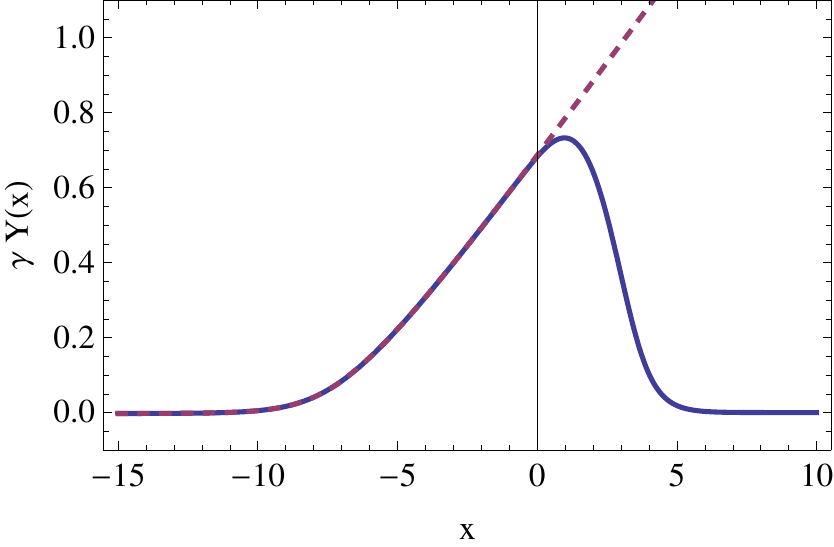}
\includegraphics[width=0.45\columnwidth]{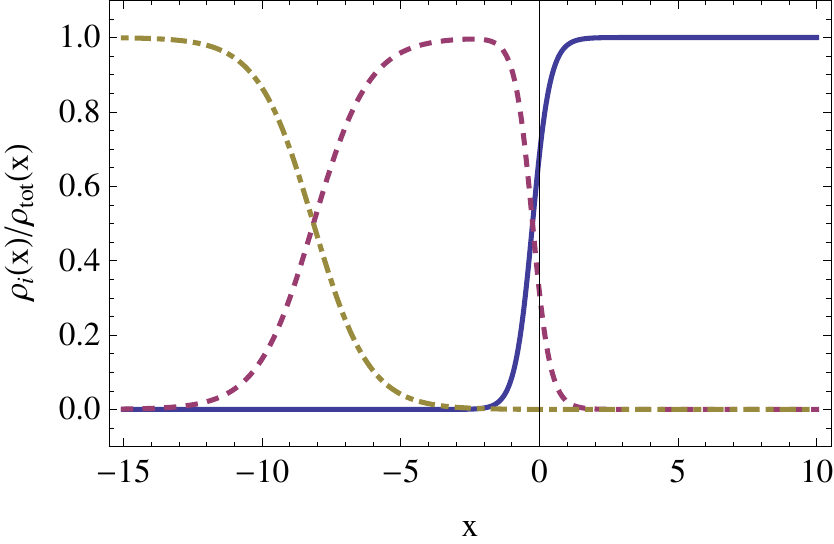}
\includegraphics[width=0.45\columnwidth]{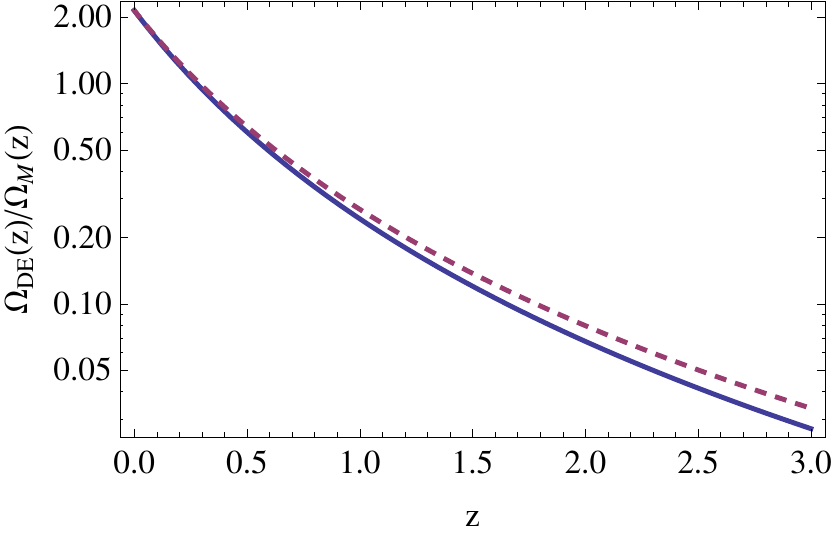}
\end{minipage}
\caption{Upper panels: the functions  $U(x)$   and $\gamma Y(x)$ from the numerical integration  of the exact equations (blue solid lines), and the corresponding perturbative solutions (dashed red);  we use
$\gamma =0.050255$. 
Lower left panel: the energy fractions  $\Omega_i=\rho_i(x)/\rho_{c}(x)$ for  $i=$~R (green, dot-dashed) $i=$~M (red, dashed) and $i=$~DE (blue solid line).   Lower right panel: the ratio $\ode(z)/\oma(z)$, shown as a function of the redshift  $z=e^{-x}-1$, in our non-local model (blue solid line) and in $\Lambda$CDM (red dashed line). \label{fig:YXU}}
\end{center}
\end{figure}

The result of the numerical integration of \eqst{sy1R}{syzR} is shown in the upper panels of
Fig.~\ref{fig:YXU} (blue solid lines).  
The red dashed lines give the corresponding perturbative solutions, that could be obtained analytically by matching the solution (\ref{pertU}), (\ref{pertY})  across the RD-MD transition or, more simply,
directly by numerical integration of \eqs{sy1R}{sy3R}, setting $\gamma=0$ in \eq{syzR}.

The behavior of $\g Y(x)$ is particularly interesting, since  $\g Y(x)$ is equal to the dark energy density $\rde(x)$ (normalized to $\rho_0$, see \eq{defrhode}). In the RD phase it remains zero, while in the MD phase it begins to grow according to the perturbative solution, and finally it becomes large and begins to dominate near the present epoch. It then decreases and goes to zero in the future, roughly as $a^{-3/2}=e^{-3x/2}$. Even if it goes to zero, 
at large $x$ this dark energy density still remains the dominant component, since it only decreases approximately as $a^{-3/2}$, while the matter density decreases as $a^{-3}$. In the lower left panel of Fig.~\ref{fig:YXU} we show  the energy fractions
$\Omega_i (t)=\rho_i(t)/\rho_c(t)$ for $i=$ radiation, matter and dark energy.
In the the lower right panel we show the ratio $\ode(z)/\oma(z)$ as a function of the
the redshift  $z=e^{-x}-1$ (blue solid line) and we compare it with the same ratio in 
$\Lambda$CDM (red dashed line). For instance,
at the value $z=1.7$ relevant for supernovae, this ratio is
0.094 for our model and  0.109 for $\Lambda$CDM. 

The fact that this model can generate a sizable DE density today, even starting from a solution that vanishes in RD,  is already a non-trivial result. Furthermore, having fixed the mass $m$ (or, equivalently, $\gamma$) from the condition $\ode=1-\oma-\ora$, we have no more free parameters and the time evolution of $\ode(x)$ is uniquely fixed, so we get
a  prediction for  evolution the  of $\rde(x)$ with $x$. This information can be compactly summarized using   fitting functions, as we now discuss.

\subsection{Fitting functions}\label{sect:fit}

In principle the function $\rho_{\rm DE}(x)/\rho_0=\g Y(x)$ computed above by numerical integration of the differential equations, and displayed in Fig.~\ref{fig:YXU}, contains all the information on the  evolution of the DE density. However, in practice it is convenient to ``coarse grain" the information contained in Fig.~\ref{fig:YXU}, expressing it in terms of a fitting function that contains just  a few parameters that can be directly compared to observations. Since the function $\ode(x)$  is negligible in the early Universe (as we see from the lower left panel in Fig.~\ref{fig:YXU}), it is actually sufficient to find a parametrization that fits it well in the recent cosmological epoch, where it start to become important. In general, the most appropriate fitting function and the corresponding best-fit parameters will depend on the range of values of $x=\ln a$ that we consider. It is useful to distinguish different case, also to have an idea of the stability of the fit.

\begin{figure}[t]
\begin{center}
\begin{minipage}{1.\linewidth}
\centering
\includegraphics[width=0.45\columnwidth]{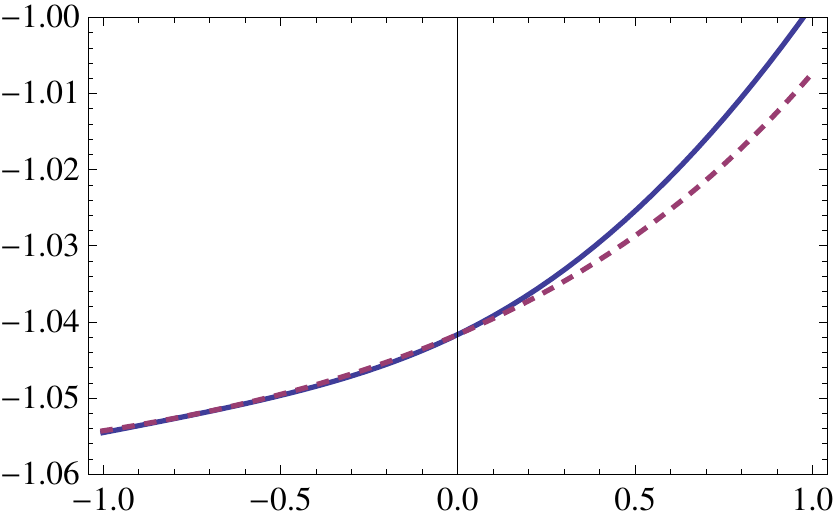}
\includegraphics[width=0.45\columnwidth]{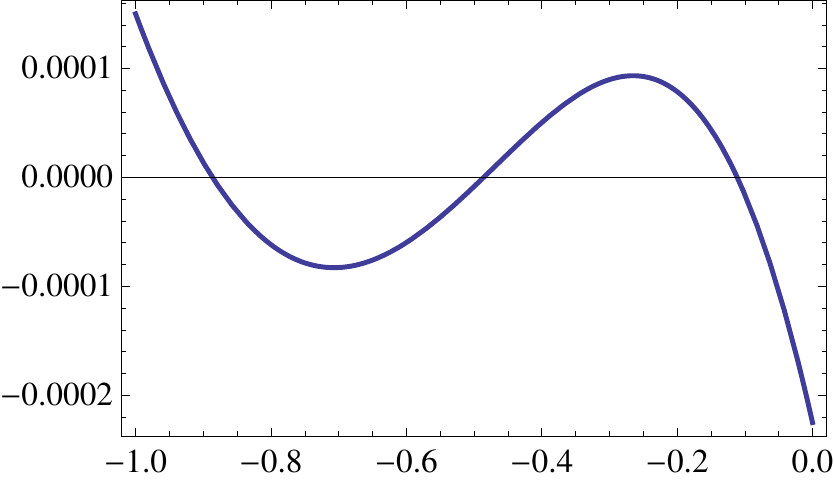}
\end{minipage}
\caption{Left: the numerical values of $w_{\rm DE}(x)$ (blue solid line) compared to the   function 
$w_{\rm DE}(a)= w_0+(1-a) w_a$ with
$w_0 = -1.0420$,  $w_a = -0.0199$
(red dashed line), in the region $-1<x<1$.
Right: the value of $\Delta w/w$, in the region $-1<x<0$. \label{fig:fitDeltaw}
}
\end{center}
\end{figure}

\vspace{3mm}
\noindent
{\bf 1.} We first consider  the   region $-1<x<0$, corresponding to redshifts $0<z\,\lsim\, 1.72$. We use the standard fitting function~\cite{Chevallier:2000qy,Linder:2002et}
\be\label{fitCP}
w_{\rm DE}(a)= w_0+(1-a) w_a\, ,
\ee  
where $a=e^x$. We define $\Delta w$ as the difference between the value of the numerical expression and this  fitting function,
and we minimize with respect to $w_0$ and $w_a$ the quantity
\be\label{chi2a}
\chi^2=\int_{-1}^0dx \, (\Delta w)^2(x)\, .
\ee
We find that the  best-fit values are
$w_0 = -1.0420$,  $w_a = -0.0199$. In the left panel of  Fig.~\ref{fig:fitDeltaw} we show the function $w_{\rm DE}(x)$ determined numerically (blue solid line) and the fitting function (\ref{fitCP}) with these best-fit values (red, dashed).  For later purposes, we also show in this figure these functions in the region $0<x<1$. We  see  that this fitting function is no longer accurate for $x>0$, which however corresponds to the future and it is therefore not relevant for the comparison with observations.
In contrast, in the region $-1\leq x<0$
the  relative error between the numerical result and the fitting function, shown of the right panel,  is at the level  $|\Delta w/w|\leq 2\times 10^{-4}$, so in this region this fitting function should be quite accurate for most purposes. 

To assess the robustness of these best-fit values under changes of the cosmological parameters we have repeated the numerical integration changing $\oma$, readjusting the  mass $m$  so that $\oma+\ora+\gamma Y(0)=1$, 
and repeating the fitting procedure. We change $\oma$ in the interval
$[0.030,0.033]$, which corresponds to the 68\% limits of Planck+WP \cite{Ade:2013zuv}.
For $\oma=0.030$, minimizing with respect to $w_0,w_a$  we get 
$w_0 = -1.0420$, $w_a = -0.0207$, while for $\oma=0.033$ we get 
$w_0 = -1.0421$, $w_a = -0.0193$. Thus, at the level of accuracy of the first three digits, the predictions of the model for $w_0$ and $w_a$ are unaffected.
We also compared with a fit of the form 
\be\label{fit2}
w_{\rm DE}(a)= w_0+(1-a^q) w_a\, ,
\ee
again restricting for the moment to the region $-1< x<0$.
Taking also $q$ as a free fitting parameter 
gives the best-fit values $w_0 = -1.0420$,  $w_a = -0.0194$ and
$q= 1.039$, but the improvement in the minimization of the $\chi^2$ is practically irrelevant, so the introduction of $q$ as a new fitting parameter  in this case is not justified.

In conclusion, in the region $-1<x<0$ the dark energy EOS is very well fitted by
\eq{fitCP},
with the best-fit values
\be\label{predw0wa}
w_0 = -1.042\, ,  \qquad w_a = -0.020\, ,
\ee
where we quoted the number of digits which is stable under changes in $\oma$ in the interval
$\oma\in [0.030,0.033]$.
For comparison, 
the  observational limit from    Planck+WP+BAO in the
$(w_0,w_a)$ plane are, at  95\% c.l. 
$w_0=-1.04^{+0.72}_{-0.69}$ and $w_a<1.32$ \cite{Ade:2013zuv}.
Actually, since our prediction for $w_a$ is such that $|w_a|\ll 1$, it  is meaningful to compare directly with
the   result of ref.~\cite{Ade:2013zuv} for a constant $w_{\rm DE}$, which is much more stringent. The result obtained combining Planck+WP+SNLS is 
$w_{\rm DE}=-1.13^{+0.13}_{-0.14}$
while Planck+WP+Union2.1 gives $w_{\rm DE}=-1.09\pm 0.17$.
The prediction  given in \eq{predw0wa} is therefore
consistent with the Planck result, and on the phantom side.\footnote{Of course, a full comparison with the Planck data also requires the computation of the cosmological perturbations in our model. Work on this is in progress.} 

The fact that the EOS parameter is on the phantom side is generically a consequence of the fact that in our model the DE density starts from zero in RD and then grows during MD. Thus, in this regime  $\rde >0$ and $\dot{\rho}_{\rm DE}>0$, and then \eq{defwDE} implies $(1+w_{\rm DE})<0$.

\vspace{3mm}

{\bf 2.} The region $-3<x<0$.
At $x=-1$ 
(which corresponds to a redshift $z\simeq 1.72$) we have $\ode(x)/\oma(x)\simeq 0.09$, which is small but not completely negligible, and
depending on the type of cosmological observations that one might  wish to use for testing the model, it can  be useful to have a fitting function that works accurately down to  lower values of $x$, e.g. down to 
the value $x=-3$ (which corresponds to a redshift $z\simeq 19$), where $\ode(x)/\oma(x)\simeq 1.5\times 10^{-4}$. At even more negative values of $x$ the effect of dark energy becomes even smaller, and in most situations a more accurate parametrization will probably not be needed. 

\begin{figure}[t]
\begin{center}
\begin{minipage}{1.\linewidth}
\centering
\includegraphics[width=0.45\columnwidth]{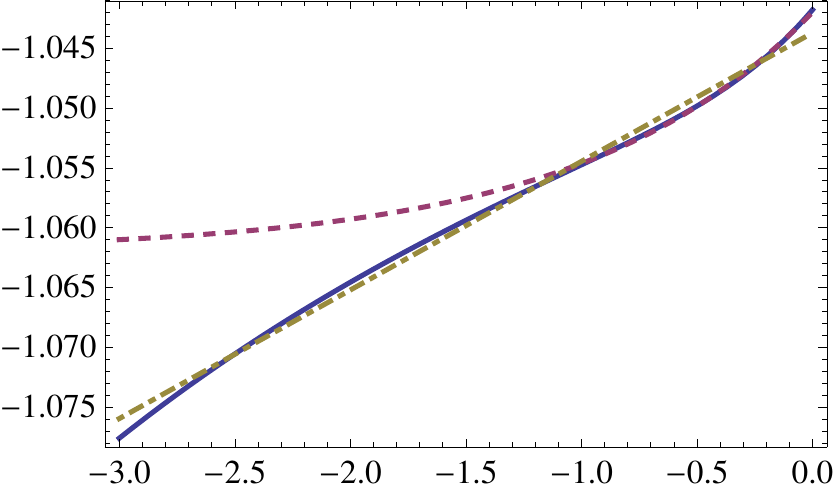}
\includegraphics[width=0.45\columnwidth]{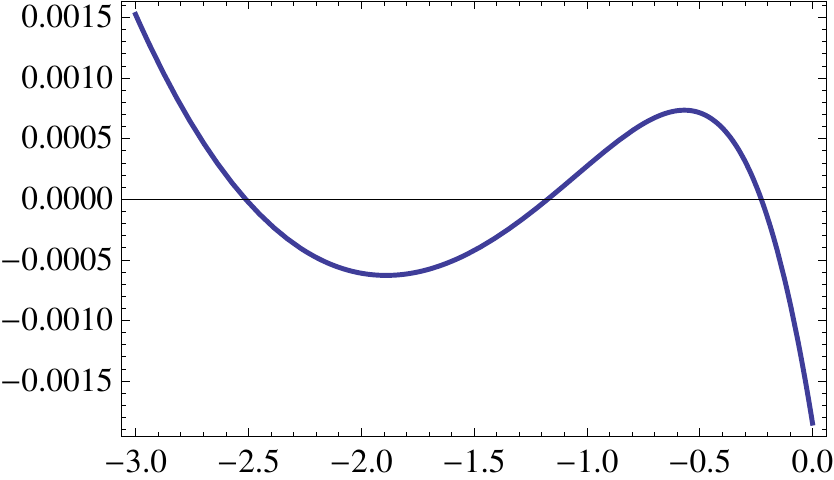}
\end{minipage}
\caption{Left panel: the EOS parameter $w_{\rm DE}(x)$ in the region $-3<x<0$ from the numerical integration (blue solid line), compared to the function (\ref{fitCP}) 
with $w_0 = -1.0420$, $w_a = -0.020$, 
(red dashed line) and to the function
(\ref{fit3}) with $w_0 = -1.0437$, $\bar{w}_a=- 0.0107$ (green dot-dashed line). Right panel: the relative error $\Delta w/w$ from the fit
(\ref{fit3}).
\label{fig:wfitx3}}
\end{center}
\end{figure}

The  fitting function (\ref{fitCP}) works quite well, as we have seen, for $-1<x<0$, but goes astray for $x<-1$, as we can see from the left panel in  Fig.~\ref{fig:wfitx3}. In contrast, a  good fitting function over the whole range $-3<x<0$ is given by
\be\label{fit3}
w_{\rm DE}(a)= w_0-\bar{w}_a\ln a\, .
\ee
Minimizing $\chi^2=\int_{-3}^0dx \, (\Delta w)^2$ with respect to  $w_0$ and $\bar{w}_a$ we get $w_0 = -1.0436$ and $\bar{w}_a= -0.0108$.
The corresponding curve 
is shown as the green dot-dashed line in the left panel of Fig.~\ref{fig:wfitx3}. It  reproduces the values from the numerical integration to a relative accuracy $|\Delta w/w|\leq 1.5\times 10^{-3}$ over the whole interval
$-3<x<0$, as we can see from the  right panel in  Fig.~\ref{fig:wfitx3}. However, 
comparing with the right panel in Fig.~\ref{fig:fitDeltaw} we see that in the region $-1<x<0$ this fit is much less accurate than the fit (\ref{fit2}).

\vspace{3mm}

{\bf 3.} Finally, we  fit $w_{\rm DE}(x)$ in the region $-1<x<1$. Of course this is to some extent academic, since only the region $x\leq 0$, i.e. our past,  is relevant for comparison with observations, but this  exercise is still instructive to get a general understanding of how the fitting function can depend on the range considered. In this case, we see from the left panel of Fig.~\ref{fig:fitDeltaw}
that the standard fit (\ref{fitCP}) is no longer  accurate, and at $x>0$ the correct fitting function should rise faster. In this  region a significantly better fit is in fact 
obtained using \eq{fit2}.
We compute the best-fit values by varying the three parameters $(w_0,w_a,q)$ so to minimize 
$\chi^2=\int_{-1}^1dx \, (\Delta w)^2$.
This gives
$w_0=-1.0414$, $w_a=-0.0189$ and  $q=1.2048$. This fit reproduces the values of the EOS obtained from the numerical solution to a relative accuracy $\Delta w/w\leq 4\times 10^{-4}$ over the range $-1<x<1$. 

\vspace{3mm}
The corresponding analytic expressions for $\rde(x)$ are obtained as usual from
energy-momentum conservation $\pa_t\rho_{\rm DE}+3[1+w_{\rm DE}(x)]H\rho_{\rm DE}=0$,  which integrates to
\be
\rde(x)=\rde(0) \exp\left\{-3\int_0^xdx' [1+w_{\rm DE}(x')]\right\}\, .
\ee
In the region $-1<x<0$ the fit (\ref{fit2}) gives
\be
\rde(x)=\rde(0) e^{-3(1+w_0)x -3w_a[x-(e^x-1)]}\, .
\ee
Note that, for $|x|\ll 1$, $\rde(x)\simeq\rde(0) e^{-3(1+w_0)x +(3/2)w_ax^2}$ and  the term 
$O(x^2)$ can be neglected, giving back the usual behavior $\rde(x)\simeq\rde(0) e^{-3(1+w_0)x}=\rde(0) a^{-3(1+w_0)}$. In the region
$-3<x<-1$ we use \eq{fit3} and we find
\bees
\rde(x)&=&\rde(0) e^{-3(1+w_0)x +(3/2)\bar{w}_ax^2}\nn\\
&=&\rde(0) e^{-3(1+w_0) (x-w_1x^2)}\, ,
\ees
where $w_1=\bar{w}_a/[2(1+w_0)]\simeq 0.1229$.

\section{A more general class of  models}\label{sect:attr}

\subsection{Freedom in the definition of the non-local operators}

The cosmological model discussed above makes use of a specific  definition of
the $\iBox_{\rm ret}$ and ${\cal D}^{-1}_{\rm ret}$ operators, given in 
\eqs{Udef}{S0Dret}. More generally, we could study the evolution  (starting again from RD) of a non-local model in which the $\iBox_{\rm ret}$ and 
${\cal D}^{-1}_{\rm ret}$ operators, applied to a function $F(t)$, are defined by
\bees
( \iBox_{\rm ret} F)(t) &\equiv& f(t)
+\int_{t_*}^{t} dt'\, \frac{1}{a^d(t')}
\int_{t_*}^{t'}dt''\, a^d(t'') F(t'')\, ,\label{Udef2}\\
({\cal D}^{-1}_{\rm ret} F)(t)&\equiv&g(t)+\int_{t_*}^{\infty} dt'\,  D_{\rm ret}(t;t') F(t')\, ,\label{S0Dret2}
\ees
where $f(t)$ is a given solution of $\Box  f=0$ and 
$g(t)$ is a given solution of ${\cal D}  g=0$. To motivate the introduction of these homogeneous solutions, consider 
 a cosmological model that starts from an earlier phase (for instance an inflationary phase)  followed by RD and then MD. In this case  it could be more natural to define $\iBox$ setting to zero the homogeneous solution at the beginning of the inflationary era, that we denote as $t=t_i$
\be\label{iBoxDW2}
(\iBox_{\rm ret}  R)(t)=-\int_{t_i}^{t} dt'\, \frac{1}{a^d(t')}
\int_{t_i}^{t'}dt''\, a^d(t'') R(t'')\, .
\ee
If we denote by $t_*$ the value of cosmic time when the inflationary epoch ends and RD starts (so $t_*>t_i$), and we compute  the value of $(\iBox_{\rm ret}  R)(t)$
during RD   using  the definition (\ref{iBoxDW2}), we have
\be
(\iBox_{\rm ret}  R)(t)=-\int_{t_i}^{t_*} dt'\, \frac{1}{a^d(t')}
\int_{t_i}^{t'}dt''\, a^d(t'') R(t'')
-\int_{t_*}^{t} dt'\, \frac{1}{a^d(t')}
\int_{t_i}^{t'}dt''\, a^d(t'') R(t'')\, ,
\ee
where we have  split $\int_{t_i}^{t} dt'=\int_{t_i}^{t_*} dt'
+\int_{t_*}^{t} dt'$.
The first integral  is just a number, 
\be
c_0\equiv -\int_{t_i}^{t_*} dt'\, \frac{1}{a^d(t')}
\int_{t_i}^{t'}dt''\, a^d(t'') R(t'')\, .
\ee
In the second integral, in contrast, $t'>t_*$, and we can use the fact that $R(t'')=0$  in RD, i.e. for $t''>t_*$, so $\int_{t_i}^{t'}dt''$ can be replaced by
$\int_{t_i}^{t_*}dt''$. Then we find that during RD, rather than having $(\iBox_{\rm ret}  R)(t)=0$ as with the definition (\ref{iBoxDW}), we now have
\be\label{c0c1iB}
(\iBox_{\rm ret}  R)(t)=c_0+c_1 f_1(t) \, ,
\ee
where 
$c_1=-\int_{t_i}^{t_*}dt''\, a^d(t'') R(t'')$ and
\be
f_1(t)=\int_{t_*}^t dt' \frac{1}{a^d(t')}\, .
\ee
Observe that both the constant $c_0$ and the function $f_1$ are solutions of the homogeneous equation $\Box f=0$, as it is clear writing $\Box =-a^{-d}\pa_0(a^d\pa_0)$. Thus, in RD this definition of $\iBox$ includes a given homogeneous solution.

The same point is also easily understood in terms of the perturbative solutions for $U$ and $Y$ given in \eqs{pertU}{pertY}. If for instance we define the non-local operators so that $u_0=u_1=a_0=a_1=0$  in the inflationary perturbative solution, during the inflationary phase we have 
\bees
U(x)&=&\frac{6(2+\zeta_0^{\rm infl})}{3+\zeta_0^{\rm infl}}x\, ,\label{Uinfl}\\
Y(x)&=&-\frac{2(2+\zeta_0^{\rm infl})\zeta_0^{\rm infl}}
{(3+\zeta_0^{\rm infl})(1+\zeta_0^{\rm infl})}
+\frac{6(2+\zeta_0^{\rm infl})}{3+\zeta_0^{\rm infl}} x
\ees
where $\zeta_0^{\rm infl}$ is the constant value of $\zeta(x)$ during inflation.
At the inflation-RD transition,  this solution will  smoothly match to a perturbative RD solution, obtained setting $\zeta_0=-2$ in \eqs{pertU}{pertY}, i.e.
\bees
U(x)&=&u_0^{\rm R}+u_1^{\rm R} e^{-x}\, ,\label{UxRD}\\
Y(x)&=&u_0^{\rm R}+a_1^{\rm R} e^{\a^{\rm R}_{+} x} +
a_2^{\rm R} e^{ \a^{\rm R}_{-}x}\, .
\ees
where $\a_{\pm}^{\rm R}=(1/2)(-5\pm\sqrt{13})$.
The value of the coefficients $u_0^{\rm R},u_1^{\rm R},a_1^{\rm R},a_2^{\rm R}$ can be determined analytically imposing the continuity of the functions and of their derivatives at the transition, and will be non-zero. 

\begin{figure}[t]
\begin{center}
\begin{minipage}{1.\linewidth}
\centering
\includegraphics[width=0.45\columnwidth]{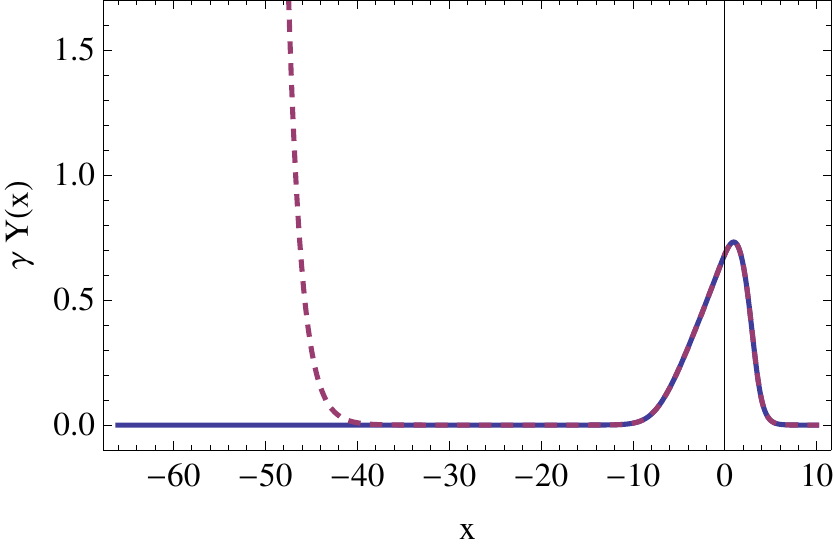}
\includegraphics[width=0.45\columnwidth]{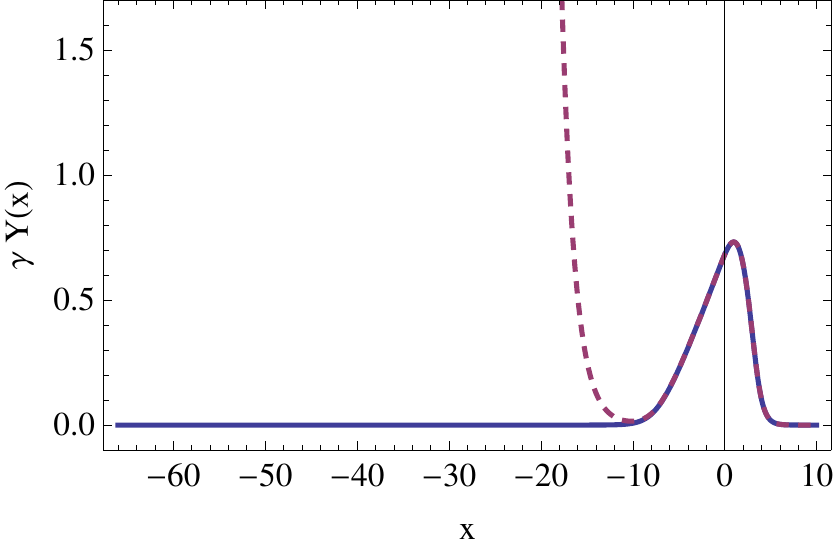}
\end{minipage}
\caption{Left panel: $\gamma Y(x)$, choosing the initial conditions on the perturbative solution (\ref{pertU},\ref{pertY})
with $u_0=u_1=a_1=a_2=0=0$ (blue solid line) compared to the solution obtained setting, at $x_{\rm in}=-66$, $Y(x_{\rm in})=10^7$ and   
$Y'(x_{\rm in})=\a_{+}Y(x_{\rm in})$ (red dashed line).
Right panel: the same with $Y(x_{\rm in})=10^{17}$.
\label{fig:Y10E}}
\end{center}
\end{figure}

The above discussion shows that, in general, even if we are interested only in the cosmological evolution starting from RD, we should in general use the definitions
of $\iBox_{\rm ret}$ and ${\cal D}^{-1}_{\rm ret}$
given in \eqs{Udef2}{S0Dret2}, allowing in general for a given homogeneous solution both in $\iBox_{\rm ret}$ and in ${\cal D}^{-1}_{\rm ret}$,
i.e.  in FRW we can in general define
\bees
U(t)&\equiv & -\iBox_{\rm ret} R \equiv U_{\rm hom}(t)
+\int_{t_*}^{t} dt'\, \frac{1}{a^d(t')}
\int_{t_*}^{t'}dt''\, a^d(t'') R(t'')\, ,\label{Udef2b}\\
S_0(t)&\equiv&{\cal D}^{-1}_{\rm ret} \dot{U}\equiv S_{0,\rm hom}(t)+\int_{t_*}^{\infty} dt'\,  D_{\rm ret}(t;t') \dot{U}(t')\, ,\label{S0Dret2b}
\ees
where $U_{\rm hom}(t)$ and $S_{0,\rm hom}(t)$ are given solution of the homogeneous equation.
These are determined 
once we assign the values of  $U,U',Y$ and $Y'$ at some initial time $x_{\rm in}$.
In turn, the set of values $\{ U(x_{\rm in}),U'(x_{\rm in}),Y(x_{\rm in}),Y'(x_{\rm in})\}$ 
can be rewritten as the set of values taken by the perturbative solutions, $\{U_{\rm pert}(x_{\rm in}), 
U'_{\rm pert}(x_{\rm in}), Y_{\rm pert}(x_{\rm in}), Y'_{\rm pert}(x_{\rm in})\}$, for a suitable choice of
the parameters $u_0,u_1,a_1, a_2$
that appears in \eqs{pertU}{pertY}. 
The advantage of using the set $\{u_0,u_1,a_1, a_2\}$ to parametrize the space of initial conditions is that in the early phase of the evolution, when we  are still deep in the RD phase, the modes proportionals to $u_0,u_1,a_1, a_2$ evolve independently, according to \eqs{pertU}{pertY}. In particular, in RD $u_1,a_1$ and $a_2$ are associated to exponentially decaying modes, so (in the space of solutions of the local model) along these directions of the parameter space the 
solution with initial conditions $u_1=a_1=a_2=0$
is an attractor. 
Thus, along these directions even relatively large initial deviations of $Y(x)$ from the unperturbed solution can be reabsorbed by the evolution. 
This is illustrated in  Fig.~\ref{fig:Y10E}, where we start at $x_{\rm in}=-66$
with a very large initial value of $Y(x_{\rm in})$ ($10^7$ on the left panel and $10^{17}$ on the right panel), and with the initial value $Y'(x_{\rm in})$ chosen equal to $\a_{+}Y(x_{\rm in})$, so that at the initial time we have  excited the mode $e^{\a_{+}x}$ 
in \eq{pertY}. However, since $\a_{+}<0$ during RD, this mode decays exponentially, and we see that even very large initial values are reabsorbed in the solution before dark energy start to be relevant (observe that, even with these large initial values, $\gamma Y(x)$ is always completely negligible with respect to the  radiation densiy in the early Universe). The mode  $e^{\a_{-}x}$ decays even faster, since  also $\a_{-}<0$, and
$|\a_{-}| >|\a_{+}|$.

The situation is different for $u_0$, which corresponds to a marginally stable mode.
In conclusion, if we discard the exponentially decaying modes associated to 
$\{u_1,a_1, a_2\}$, the only interesting generalization of our non-local model is obtained defining, in RD,
\be\label{gener}
 -\iBox_{\rm ret} R \equiv u_0
+\int_{t_*}^{t} dt'\, \frac{1}{a^d(t')}
\int_{t_*}^{t'}dt''\, a^d(t'') R(t'')\, ,
\ee
while in the definition (\ref{S0Dret2}) we can still keep $S_{0,\rm hom}(t)=0$.
We now explore the physical meaning and the cosmological consequences of this modification.

\subsection{Modification of $\iBox$ and the cosmological constant}

Quite interestingly, the introduction of $u_0$ is equivalent to introducing an explicit cosmological constant term in the non-local model. Indeed, let us write
\be\label{newold}
-\iBox_{\rm new} R \equiv u_0-\iBox_{\rm old}R\, ,
\ee
where $\iBox_{\rm new} R$ is the new definition of the retarded $\iBox$ operator given in \eq{gener} while $\iBox_{\rm old} R$ is our ``old" definition
(\ref{iBoxDW}). The model which uses the new definition is governed by the equation
\be\label{new1}
\Gmn -m^2\frac{d-1}{2d}\(\gmn \iBox_{\rm new}  R\)^{\rm T}=8\pi G\,\Tmn\, ,
\ee
which, using \eq{newold}, becomes
\be\label{new2}
\Gmn -m^2\frac{d-1}{2d}\(\gmn \iBox_{\rm old}  R\)^{\rm T}=8\pi G\,\Tmn-\Lambda\gmn\, ,
\ee
with  $\Lambda =[(d-1)/2d] m^2u_0$. We have therefore re-introduced a cosmological constant! Writing $\rlam =\Lambda/(8\pi G)$ and
$\ola =\rlam/\rho_0$, 
in $d=3$ we get
\be
\ola =\frac{m^2u_0}{9H_0^2}=\gamma u_0\, .
\ee
The result is quite interesting because it shows that, once we discard the modes that are exponentially decaying during RD, the whole freedom in the definition of
the non-local operators $\iBox_{\rm ret}$ and ${\cal D}^{-1}_{\rm ret}$ boils down to the possibility of introducing an explicit cosmological constant in the equations, with a  values determined by $\gamma$ and by the initial conditions on the auxiliary field $U$.
In the next section we will study the cosmological evolution of this more general class of models.

\subsection{Cosmological evolution for $u_0\neq 0$}

\begin{figure}[t]
\begin{center}
\begin{minipage}{1.\linewidth}
\centering
\includegraphics[width=0.45\columnwidth]{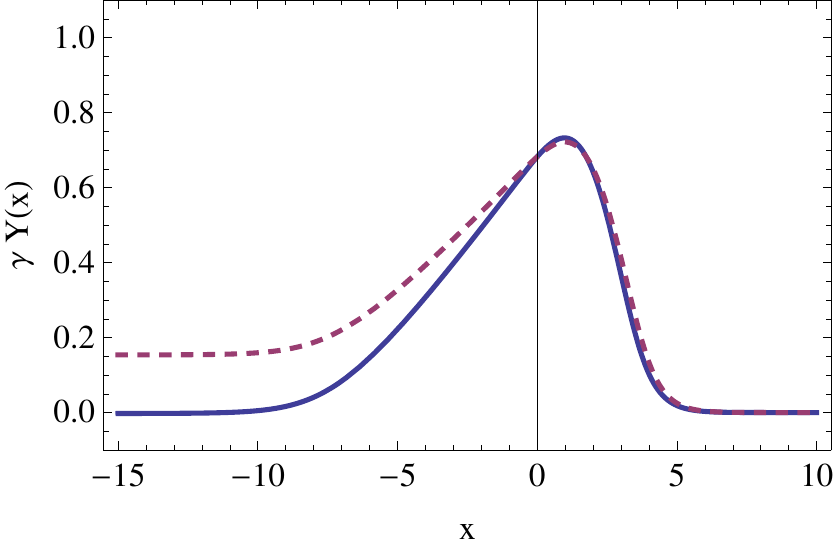}
\includegraphics[width=0.45\columnwidth]{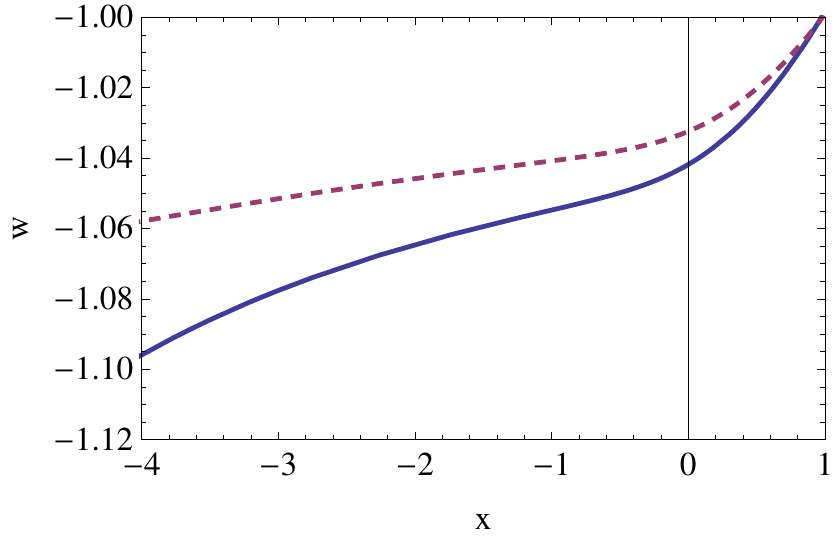}
\end{minipage}
\caption{Left panel:  $\gamma Y(x)$, choosing the initial conditions on the perturbative solution (\ref{pertU},\ref{pertY})
with $u_0=0$ (blue solid line) and with $u_0=4$ (red dashed line). In both cases we adjusts $\gamma$ so to maintain  fixed $\oma=0.3175$. 
Right panel: the EOS  parameter for $u_0=0$  (blue solid line) and  $u_0=4$ (red dashed line).
\label{fig:c4_fixe_OM}}
\end{center}
\end{figure}

\begin{figure}[t]
\begin{center}
\begin{minipage}{1.\linewidth}
\centering
\includegraphics[width=0.45\columnwidth]{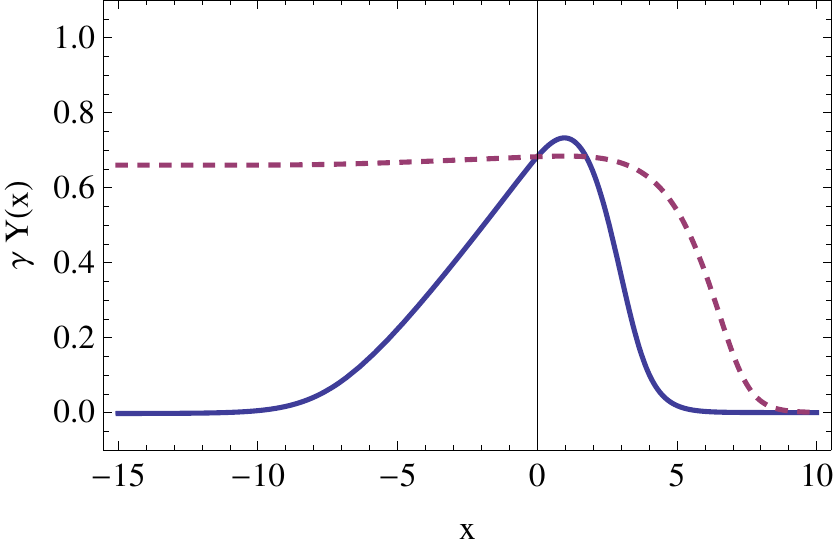}
\includegraphics[width=0.45\columnwidth]{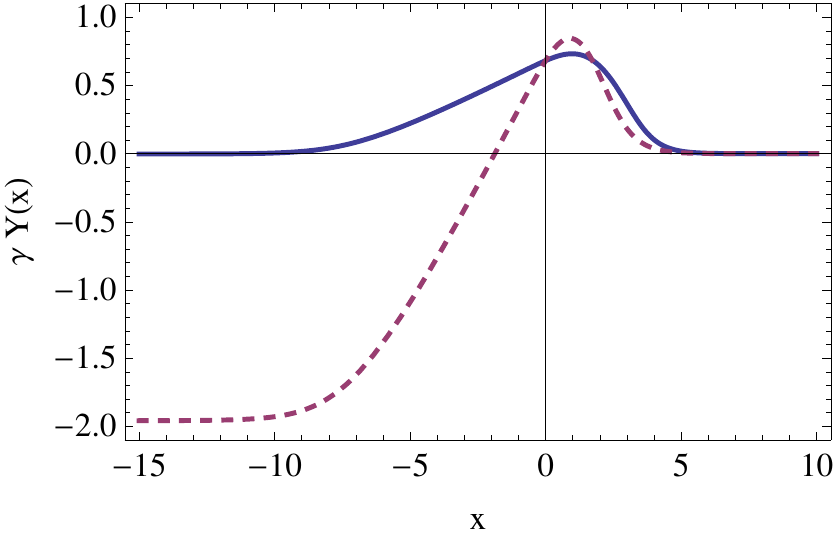}
\end{minipage}
\caption{Left: $\gamma Y(x)$, choosing the initial conditions on the perturbative solution 
with $u_0=0$ (blue solid line) and with $u_0=400$ (red dashed line).
Right: the same for $u_0=0$ (blue solid line) and  $u_0=-10$ (red dashed line).
\label{fig:u0400}}
\end{center}
\end{figure}

The evolution  equations, with the definition (\ref{gener}) of the $\iBox_{\rm ret}$ operator, are still given by \eqst{hLCDM}{syzR}, except that now the initial conditions
on $U$ and $Y$ in RD  are $U(x_{\rm in})=Y(x_{\rm in})=u_0$ and
$U'(x_{\rm in})=Y'(x_{\rm in})=0$, 
as we see from \eqs{pertU}{pertY} with $\zeta_0=-2$. 
The effect of using a value of $u_0$  of order one
is illustrated in the left panel of Fig.~\ref{fig:c4_fixe_OM}, where we compare the solution of the previous section, found  choosing the initial conditions on the perturbative solution (\ref{pertU},\ref{pertY})
with $u_0=0$, to the solution found setting $u_0=4$.
In both cases we adjust $\gamma$ so to maintain  fixed $\oma=0.3175$, which is obtained using  $\gamma =0.050255$ for $u_0=0$ and 
$\gamma =0.038930$ for $u_0=4$.\footnote{Let us stress again that, even if from the point of view of the local formulation $u_0$ enters  through the initial conditions, at the level of the original non-local formulation each value of $u_0$ defines a {\em different} theory. Each theory will be characterized by its own value of $\gamma$, required in order to get $\oma=0.3175$ today.} We see that, when $u_0>0$, the dark energy density has a constant component which is non-vanishing even in RD.
The respective EOS parameters are shown in the right panel. The effect of introducing a positive $u_0$ is to increase $w_0$ from the value $-1.042$ that it has for $u_0=0$, toward a value closer   to $-1$, but still on the phantom side (for $u_0=4$, we get $w_0=-1.032$). This is clearly understood from the fact that the introduction of $u_0$ is formally equivalent to introducing a cosmological constant, with
$\ola=\gamma u_0$, on top of which evolves a dynamical dark energy, with the sum of these two components still constrained to take the value $0.68$ today. Therefore the value of $w_0$ is  shifted toward the value $-1$ corresponding to a cosmological constant. 
This effect can be seen even more clearly taking a much larger value of $u_0$, e.g.
$u_0=400$. In this case we must choose 
$\gamma=0.00165$ in order to keep  fixed $\oma=0.3175$. The result is shown in the left panel of Fig.~\ref{fig:u0400}. In this case  we find $w_0\simeq -1.001$. 
It is clear that, as $u_0\ra+\infty$, the model approaches more and more 
$\Lambda$CDM. In fact, in order to keep the contribution $\ola=\gamma u_0$ at a value smaller or equal than $0.68$, as we send $u_0\ra\infty$ we must tune $\gamma\ra 0$; correspondingly, the dynamical contribution $\gamma Y(x)$ to the DE goes to zero, and in the limit $u_0\ra\infty$ we remain with a $\Lambda$CDM model with 
$\gamma u_0$ kept fixed at the value 0.68. At $u_0\gsim 100$, we find that the numerical results for $w_0$ and $w_a$ are well fitted by
\be\label{w0u0}
w_0\simeq -1 -\frac{A}{u_0}\, ,\qquad
w_a\simeq-\frac{B}{u_0}
\ee
with $A\simeq 0.5$, $B\simeq 0.1$, 
and therefore 
\be\label{waw0largeu}
w_a\simeq \frac{B}{A}(1+w_0)\simeq 0.2\,  (1+w_0)\, .
\ee
So, in the more general class of model parametrized by $u_0$, the   prediction (\ref{predw0wa}) changes. 
Nevertheless, we see that even in this more general class of models the EOS parameter  today is always on the phantom side, and while $u_0$ spans the whole range $u_0\in [0,\infty)$, the prediction for $w_0$ remains in the rather narrow range $[-1.042,-1)$. Furthermore, we have a relation between $w_0$ and $w_a$ in which $u_0$ is eliminated, and which therefore remains as a pure prediction of the model.

It is also interesting to explore the region $u_0<0$. In this case we are effectively adding a negative value of $\ola$. The resulting evolution is given in right panel of Fig.~\ref{fig:u0400} for $u_0=-10$. In this case  we find $w_0\simeq -1.12$, $w_a\simeq -0.32$. Observe that, for $u_0<0$, the value of  $w_0$ is shifted even more toward the phantom side. Varying $u_0$ we find that for $u_0$ lower than a critical value $u_c\simeq -12$ it is no longer possible to obtain $\gamma Y(0)=0.68$. This is due to the fact that the function $Y(x)$ can only begin to rise at the beginning of the MD era, and if   at this epoch it starts from a value below a critical one, it cannot rise fast enough to attain the required value in $x=0$. This again has the effect of limiting the possible range of predictions for $w_0$ and $w_a$ in our model.

\begin{figure}[t]
\begin{center}
\begin{minipage}{1.\linewidth}
\centering
\includegraphics[width=0.45\columnwidth]{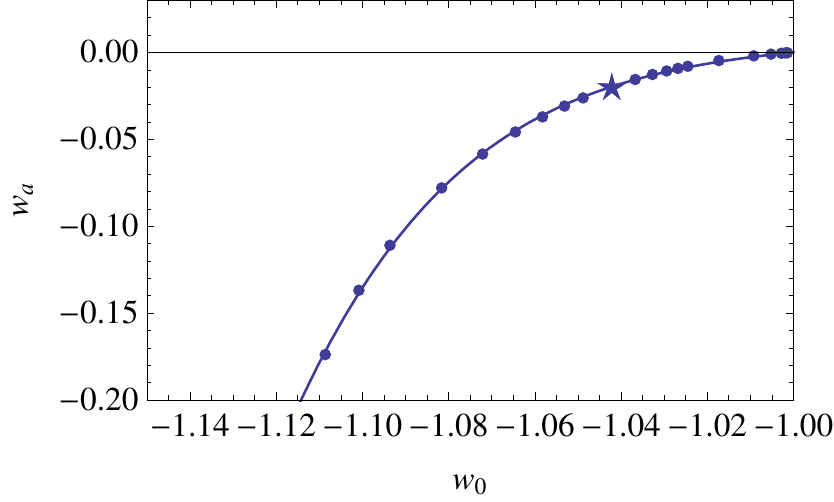}
\end{minipage}
\caption{The values of the pair $(w_0,w_a)$ obtained for different $u_0$ (dots). The star corresponds to the model with $u_0=0$.
\label{fig:Plotw0wa}}
\end{center}
\end{figure}

In Fig.~\ref{fig:Plotw0wa} we show the values of the pair $(w_0,w_a)$ obtained for different $u_0$ (as defined by  fitting to the function 
$w(a)=w_0+(1-a)w_a$ in the region $-1<x<0$). In order not to clutter the diagram, we only show a subset of the points actually computed. The star marks the position of the point  
$(w_0 = -1.042,w_a = -0.020)$ obtained for $u_0=0$. Toward its right we have
displayed the points computed for $u_0=2,4,6,8,10,20,50,100,200,300$ and $400$,
while at its left we have shown the points obtained for $u_0=-2,-3,\ldots ,-8,-8.5,-9$.
The solid line in Fig.~\ref{fig:Plotw0wa} is a fit to these data of the form 
\be\label{wafit}
w_a=a (1+w_0)+b(1+w_0)^2+c(1+w_0)^4\, ,
\ee
with $a\simeq 0.231$, $b\simeq -4.386$ and $c\simeq -684.2$.

As we move toward the value $u_c\simeq -12$, a fit of the form 
$w(a)=w_0+(1-a)w_a$ in the range $-1<x<0$ is no longer appropriate. This is due to the fact that,
when $Y(x)$ starts from negative values, it must cross the horizontal axis somewhere in order to reach a positive value at $x=0$, as in the right-panel of Fig.~\ref{fig:u0400}. At this point $Y'/Y$ diverges, and therefore also $w(x)$. If this happens within the interval $[-1,0]$, our fitting procedure is no longer appropriate. In this regime it is better to use a ``pointlike" definition $w_0=w(0)$, $w_a=-w'(0)$ (which, for values of $u_0\geq 0$  gives results consistent, at the level of three decimal figures, with those obtained fitting to the region $-1<x<0$). In this case, at the last point $u=u_c\simeq -12$ for which 
we can obtain an evolution such that $\gamma Y(0)=0.618$, we get
\be
w_0\simeq -1.33\, ,\qquad w_a\simeq -0.58\, ,
\ee
which we take as the most extreme prediction of our model. This is however just a small corner of our parameter space. As we see from Fig.~\ref{fig:Plotw0wa},
in the rest of the parameter space $w_0$ and $w_a$ vary over a much more narrow range.

\subsection{Estimating the values of $u_0$ from an earlier inflationary phase}\label{sect:infl}

In the above analysis, the value of $u_0$ during RD has been taken as a free parameter. However, we have seen that such a non-zero value is naturally generated by the evolution from a pre-existing inflationary phase, and it is interesting to try to estimate it in terms of the parameters of such a phase.
Consider then a model that starts in an earlier inflationary phase, followed by RD and MD.
As we saw in \eq{w013},
during an inflationary phase the coefficient $\a_+$ in \eq{pertY} is positive, and the corresponding homogeneous solution is unstable.  However, if the $\iBox_{\rm ret}$ operator is defined setting to zero the homogeneous solutions during the inflationary phase,   the  solutions 
$e^{\a_{\pm}x}$ are  spurious, and  are simply not solutions of the original non-local equation. Thus,  in the space of solutions of the original non-local equation,  the perturbative inhomogeneous solution during the inflationary phase is stable, even if in the space of solutions of the
differential equations of the local formulation it is unstable. The unstable direction is a spurious  solutions, which has been introduced by the localization procedure. 

As shown in \eqst{iBoxDW2}{c0c1iB}, this definition of  the $\iBox_{\rm ret}$ operator in the inflationary phase will generate a non-vanishing homogeneous solution in RD, and we  can find the corresponding values  of 
the constants $u_0^{\rm R}, u_1^{\rm R}, a_1^{\rm R}$ and $a_2^{\rm R}$. This exercise should be taken with some care, because we are assuming that the non-local massive gravity model that we are considering is valid up to the energies, such as $10^{16}$~GeV, where inflation takes place. Actually, as discussed in \cite{Maggiore:2013mea,FMMghost}, the non-local equation (\ref{modela1b2}) should be understood as an effective classical equation obtained from some form of classical or quantum smoothing in a more fundamental theory, so
the model might be modified in the UV well before such scales are reached.
Furthermore, we  must anyhow  make assumptions on the values of $U$ and $Y$ at the beginning of inflation. 
With these caveats, lets us define the non-local operators so that 
the solution (\ref{pertY}) in the inflationary phase has
$u_1=a_1=a_2=0$, and choose $u_0$ so that $U(x)=0$ at the beginning of inflation, $x=x_i$. 
Setting for simplicity $\zeta_0^{\rm infl}=0$ in \eq{Uinfl} (which corresponds to a phase of de~Sitter inflation),   \eqs{pertU}{pertY}  then give
\be
U(x)=Y(x)=4(x-x_{\rm i})
\ee
during inflation.
Denoting by   $x=x_{\rm f}$ the value of $x$ where inflation ends and RD begins,  we have 
$U(x_{\rm f})=Y(x_{\rm f})=4\Delta N$, where $\Delta N=x_{\rm f}-x_{\rm i}$ is the number of inflationary e-folds. Depending on the energy scale at which inflation takes place, the minimum number of e-folds required for a successful de~Sitter inflationary model ranges from 
$\Delta N\simeq 67$ for an inflationary scale  at $10^{16}$~GeV, to 
$\Delta N\simeq 37$ for  inflation   at the TeV. This gives a minimum value
of $U(x_{\rm f})\simeq 150$, while taking  $\Delta N=67$  we have
$U(x_{\rm f})= 268$.

Performing the matching to the analytic RD solution we find that during RD
\bees
U(x)&=&u_0^{\rm R}
-4e^{-(x-x_{\rm f})}\, ,\\
Y(x)&=&u_0^{\rm R}+c_1 e^{\a_{+}(x-x_{\rm f})}+
c_2 e^{\a_{-}(x-x_{\rm f})}\, .
\ees
where
\be\label{u0R}
u_0^{\rm R}=4(\Delta N +1)\simeq 4\Delta N\, ,
\ee
and $c_1, c_2={\cal O}(1)$.
Since in RD $\a_{\pm}<0$, all exponentials  decay, and this solution 
is quickly attracted toward the solution with $u^{\rm R}_1=a^{\rm R}_1=a^{\rm R}_2=0$. Thus, the subsequent evolution is identical to that obtained setting initial condition in the RD phase such that $u_0^{\rm R}$ is given by \eq{u0R} while
$u^{\rm R}_1=a^{\rm R}_1=a^{\rm R}_2=0$.
Since $u^{\rm R}_0\gsim 100$ we can use the fit  (\ref{w0u0}) and  (under the hypothesis specified above) we  get a  prediction for $w_0$ and $w_a$ in terms of the number of inflationary e-folds, 
\be\label{w0u0DN}
w_0\simeq -1 -\frac{1}{8\,\Delta N}\, ,\qquad
w_a\simeq-\frac{1}{40\,\Delta N}\, ,
\ee
as well as the relation (\ref{waw0largeu}) (or, more accurately, the relation
(\ref{wafit})) between $w_0$ and $w_a$. Observe that in this case $w_0$ will be very close to $-1$. For $\Delta N=67$, \eq{w0u0DN} gives
$w_0\simeq -1.002$, while for $\Delta N=37$ we get $w_0\simeq -1.003$.

\section{Conclusions}

In this paper we have analyzed the cosmological consequences of  a non-local generalization of GR that, in the far IR, involves the addition to the Einstein equations of a term proportional to the transverse parts of  $\gmn\iBox R$, 
\eq{modela1b2}. This models can be seen broadly  as a classical theory of massive gravity, in the sense that GR is deformed by the introduction of a mass parameter (although, as discussed in \cite{Maggiore:2013mea}, the graviton in this theory remains massless!).    A rather appealing feature of the $\gmn\iBox R$ model is that it is highly predictive, since, contrary to typical scalar-tensor theories,  $f(R)$ theories, 
$Rf(\iBox R)$ theories, etc., we do not have arbitrary functions of the scalar field or of the curvature that enter the model, and which are normally chosen so to have the desired cosmological behavior.
In our model in a first approximation we only have one free parameter, the mass scale $m$, or equivalently the dimensionless parameter $\gamma= m^2/(9H_0^2)$, 
which replaces the parameter $\ola$ in $\Lambda$CDM. 
We have also seen that the model can be extended adding the most general solution of the homogeneous equations $\Box U=0$ and ${\cal D}S_0=0$ in the definition of
the $\iBox$ and ${\cal D}^{-1}$ operators. In the end,  this amounts to adding to the corresponding local system a more general set of initial condition, parameterized by  four variables $u_0,u_1,a_1,a_2$. However, $u_1,a_1,a_2$ parametrize irrelevant directions in the space of solutions. It is therefore natural to consider a ``next-to-minimal" model, in which only $u_0$ is retained.
We have found that the introduction of $u_0$ is equivalent to adding a cosmological constant on top of the dynamical dark energy components.  In this sense, the existence of this marginally stable direction of parameter space is not surprising. It is clear that, given a cosmological model that produces a dynamical dark energy, we can always put on top of it the contribution of a cosmological constant.

At the level of background evolutions, these models provide quite interesting predictions:

\begin{enumerate}
\item  In the case $u_0=0$, as already shown in
\cite{Maggiore:2013mea},
 once we fix $\gamma$ so to reproduce the observed value of the DE density today, $\ode\simeq 0.68$, we have no more freedom, and we get 
a sharp prediction for the dark energy equation of state. Writing
$w(a)=w_0+(1-a)w_a$, we get 
$w_0 = -1.042$ and $w_a = -0.020$, see \eq{predw0wa}. Various aspects of this result are quite interesting. First of all it is highly non-trivial that, without any tuning, we get a value of $w_0$ so close to $-1$. Second, the result is on the phantom side, as suggested, at the $2\sigma$ level, by the Planck results \cite{Ade:2013zuv}.

\item In the models parametrized by $m$ and $u_0$ we have one more free parameter and therefore, unavoidably, less predictivity.  Nevertheless, even opening this new direction in parameter space, the predictions of the model remain quite sharp. First of all, the parameter $w_0$ always remain on the phantom side. As $u_0$ sweeps the range $u_0\in [0,\infty)$, the prediction for $w_0$ remains in the rather narrow range $[-1.042,-1)$, moving monotonically toward $-1$ as $u_0\ra +\infty$. Similarly, 
$w_a$ moves monotonically from the value that it has for $u_0=0$, $w_a = -0.020$,
toward the value $w_a=0$ for $u_0\ra\infty$. As $u_0\ra\infty$ we then approach 
the $\Lambda$CDM point $(w_0=-1, w_a=0)$. 

We have also changed $u_0$ toward negative values and found that, below a critical value $u_0\simeq -12$, it is no longer possible to obtain $\ode\simeq 0.68$ today. In the range of allowed negative values of $u_0$, $w_0$ goes even more toward the phantom side, reaching a minimum value
$w_0\simeq -1.33$, with $w_a\simeq -0.58$.

In conclusion these  model  generically predict  a value of $w_0$ on the phantom side, in the relatively narrow range $-1.32\,\lsim\ w_0 \leq -1$. Furthermore, if $w_0$ is measured with sufficient accuracy and is within this range, we can deduce from it the value of $u_0$ and therefore get a pure prediction for $w_a$. Equivalently, our model predicts a relation between the observed values of $w_0$ and $w_a$, which is displayed in Fig.~\ref{fig:Plotw0wa} and fitted in \eq{wafit}.

\end{enumerate}

The phantom value of $w_0$ that we find is quite suggestive, in view of the Planck results. The target of the Euclid mission is to reach a precision of 0.01 on $w_0$ and of 0.1 on $w_a$ \cite{Amendola:2012ys}. At this level of precision, we will have a very stringent test of the prediction given in \eq{predw0wa}, or more generally of the relation between $w_a$ and $w_0$ given in Fig.~\ref{fig:Plotw0wa}.

\vspace{5mm}

\noindent
{\bf Acknowledgments.} We thank  Luca Amendola, Ed Copeland, Yves Dirian,  Valeria Pettorino  and Christof Wetterich for useful discussions. Our work is supported by the Fonds National Suisse.

\appendix

\section{Cosmological dynamics of the $\iBox\Gmn$ model.}\label{sect:iGmn}

\subsection{Cosmological evolution equations}

In this appendix we consider the model obtained setting  $b_1=1, b_2=0$ in
\eq{modela1a2},
\be\label{final5}
\Gmn -m^2 \(\iBox_{\rm ret}\Gmn\)^{\rm T}=8\pi G\,\Tmn\, .
\ee
We now define
\be\label{defSmunu}
S_{\mu\nu}\equiv \iBox_{\rm ret}\Gmn\, ,
\ee
and we split it into its transverse and longitudinal parts, as in \eq{splitSmn}.
We can then rewrite \eq{final5} as a pair of differential equations
\bees
\Gmn -m^2 S_{\mu\nu}^{\rm T}&=&8\pi G\,\Tmn\, ,\label{pair1}\\
\Box S_{\mu\nu}&=&\Gmn\, .\label{pair2}
\ees
In the case of FRW, the procedure for extracting the transverse part from
$\Smn$ and reducing \eqs{pair1}{pair2}
to a system of ordinary differential equations  has been described  in \cite{Jaccard:2013gla}. 
In FRW, for symmetry reasons the only non-vanishing components of the tensor $\Smn$ are
$S^0_0(t)$ and $ S^i_i(t)$ (where the sum over $i$ is understood). Similarly, the only non-vanishing component  of the vector $S_{\mu}$ that enters in  \eq{splitSmn} is $S_0$. 
Using the combinations
\be\label{UV}
U=S^0_0+ S^i_i\, ,\qquad V=S^0_0-\frac{1}{d}S^i_i\, , 
\ee
as well as 
$W=-(d+1)S_0$, one finds  a system of four coupled equations for the four functions $H,U,V,W$. Specializing henceforth to $d=3$, the result is \cite{Jaccard:2013gla}
\bees
H^2+\frac{m^2}{12}\( U+3V-\dot{W}\)&=&\frac{8\pi G}{3}(\rho_M+\rho_R)\, , \label{UVW0}\\
\ddot{W}+3H\dot{W}-3H^2W &=& \dot{U}+3\dot{V}+12HV\, ,\label{UVW1}\\
\ddot{U}+3H\dot{U} &=&
6\dot{H}+12H^2\, ,\label{UVW2}\\
\ddot{V}+3H\dot{V} -8 H^2V &=&
-2\dot{H}\label{UVW3}\, ,
\ees
where, as  in sect.~\ref{sect:iBoxR},
we have taken  $\rho$ equal to the sum of the matter density $\rho_M$ and the radiation density $\rho_R$. We then define
\be\label{rhoDE}
\rho_{\rm DE}(t)\equiv \frac{m^2}{32\pi G}(\dot{W}-U-3V)
\, ,
\ee
so  \eq {UVW0} takes again the  form
\be\label {UVW0b}
H^2(t)=\frac{8\pi G}{3}\[\rho_M(t)+\rho_R(t)+\rho_{\rm DE}(t)\]\, .
\ee
We pass again to dimensionless variables as  in sect.~\ref{sect:iBoxR}, we use $x=\ln a(t)$ instead of $t$, and
we  also trade $W$ for a field $Y$ defined by
$Y=\dot{W}-U-3V=hW'-U-3V$.
Then the Friedmann equation reads
\be\label{syh}
h^2(x)=\Omega_M e^{-3x}+\Omega_R e^{-4x}+\g Y(x)\, ,\\
\ee
where now $\gamma=m^2/(12 H_0^2)$, 
and the evolution of $Y(x)$ is obtained from the coupled system of equations
\bees
&&Y''+(3-\zeta)Y'-3(1+\zeta)Y=-3U'+3(1+\zeta) U +3V'+3 (3-\zeta)V\, ,\label{sy1}\\
&&U''+(3+\zeta)U'=6(2+\zeta)\label{sy3}\, ,\\
&&V''+(3+\zeta)V'-8V=-2\zeta\, ,\label{sy4}
\ees
where  
\be
\zeta(x)\equiv\frac{h'}{h}=-\frac{1}{2}\, \,
\frac{3\Omega_M e^{-3x}+4\Omega_R e^{-4x}
-\g Y'(x) }{\Omega_M e^{-3x}+\Omega_R e^{-4x}+\g Y(x)}\label{syz}\, .
\ee
Just as with the $\iBox R$ model of sect.~\ref{sect:iBoxR}, we see from these equations that, at the level of background evolution, compared to $\Lambda$CDM  the cosmological constant term is replaced by a dark energy term  with
$\rde(x)=\rho_0\g Y(x)$
or, in terms of the dark energy fraction $\ode(x)$, 
\be\label{odeY}
\ode(x) \equiv\frac{\rde(x)}{\rho_c(x)}=\frac{\g Y(x)}{h^2(x)}\, ,
\ee
and the dynamics of $Y(x)$  is governed by the coupled system of equations (\ref{sy1})-(\ref{sy4}). 
The  effective EOS parameter of this dark energy component is again defined by
$\dot{\rho}_{\rm DE}+3(1+w_{\rm DE})H\rho_{\rm DE}=0$, which gives again \eq{wg2}.

\subsection{Perturbative solutions and instabilities}

Neglecting the contribution of $Y$ to $\zeta$ and setting $\zeta(x)\simeq \zeta_0$
the equations for $U$ and $V$ become
\bees
U''+(3+\zeta_0)U'&=&6(2+\zeta_0)\label{sy3z0}\, ,\\
V''+(3+\zeta_0)V'-8V&=&-2\zeta_0\, ,\label{sy4z0}
\ees
whose solution is (see also \cite{Modesto:2013jea})
\bees
U(x)&=&\frac{6(2+\zeta_0)}{(3+\zeta_0)}x+u_0
+u_1 e^{-(3+\zeta_0)x} \,,\label{pertUa}\\
V(x)&=& \frac{\zeta_0}{4}+v_0 e^{\b_+ x} +v_1 e^{\b_{-}x}\, ,\label{pertVa}
\ees
where 
\be
\b_{\pm}=-\frac{3+\zeta_0}{2}\pm \sqrt{\(\frac{3+\zeta_0}{2}\)^2+8}\, .
\ee
In particular, during RD, $\b_{\pm}=(1/2) (-1\pm \sqrt{33})$, and during MD
$\b_{\pm}=(1/4) (-3\pm \sqrt{137})$. The  solutions for $Y$ is obtained plugging
\eqs{pertUa}{pertVa} into \eq{sy1} (with $\zeta(x)$ replaced by $\zeta_0$) and is of the form
\bees
Y(x)&=&c_0+c_1x  +3u_0(1+\zeta_0)+ c_2 u_1e^{-(3+\zeta_0)x}
+c_3 v_0 e^{\b_+ x} +c_4 v_1 e^{\b_{-}x} \nn\\
&&+y_0  e^{\a_{+}x}+ y_1 e^{\a_{-}x}\, .\label{Ypert}
\ees
The coefficients $c_0,\ldots , c_4$ are functions of $\zeta_0$ easily obtained by direct substitution (and whose relatively cumbersome expression we will not need below). The terms proportional to $y_0,y_1$ are  the general solution of the homogeneous equation $Y''+(3-\zeta_0)Y'-3(1+\zeta_0)Y=0$, and
$\a_{\pm}=(1/2)[-3+\zeta_0\pm \sqrt{21+6\zeta_0+\zeta_0^2}]$.
The  solutions of the inhomogeneous equations obtained setting $u_0=u_1=v_0=v_1=y_0=y_1=0$
are self-consistent with the  perturbative approach, since at early times (i.e. as $x\ra-\infty$)  $Y(x)\propto x$, so its contribution to $\zeta(x)$  in \eq{syz} is indeed negligible compared to $\oma e^{-3x}$ and $\ora e^{-4x}$.  Therefore they provide a solution of the equations that gives back standard cosmology at early times.

In sect.~\ref{sect:iBoxR} we found that, for the $\iBox R$ model, the homogeneous solutions are stable (or, in the case of $u_0$, marginally stable) both in RD and MD. There is a potential instability if there is an earlier inflationary phase, which can however be avoided assigning the appropriate boundary conditions that exclude them during inflation.\footnote{Furthermore, the subsequent exponential decrease of the solution during RD and MD still allows us to obtain a sensible cosmological evolution even if there is an exponentially growing term during inflation. Observe also that, despite this exponential growth, the DE density $\rde (x)$ remains utterly negligible compared to $\rho_c(x)$ in the inflationary as well as  in the subsequent RD phases.}
In contrast, in this model the homogeneous solution for 
$V(x)$ associated to the 
mode $e^{\b_{+}x}$ is unstable both in RD and in MD, since in both regimes $\b_{+}>0$. This instability makes it impossible to obtain a convincing evolution during RD and MD. If we start the evolution from an earlier inflationary era, even setting to zero the homogeneous solution during this epoch, once we enter in RD and we match the perturbative solution during inflation with the perturbative solution during RD, the homogeneous solutions of the RD era will be generated, and will quickly lead to an instability of the system. We could start the evolution from the RD era, assigning there initial conditions that amount to setting to zero the homogeneous RD solution, but in any case the instability will show up in MD. We have indeed checked this behavior with the numerical integration of  the exact equations
(\ref{syh})--(\ref{sy4}). The instability is triggered by the exponentially growing mode of $V(x)$ but, since $V(x)$ couples to all other functions, it leads to an instability also in  the functions $U$ and $Y$ and then in the Hubble parameter $h(x)$, which leads to an early phase of accelerated expansion that screws up the standard RD and MD epochs. The conclusion is that the model with $b_1=1, b_2=0$ is not cosmologically viable, since already at the level of background evolution it cannot reproduce standard cosmology at early times.\footnote{Alternatively, one could set the graviton mass to extremely small values compared to $H_0$, so to suppress the  instability and therefore the early beginning of the acceleration era, as suggested in \cite{Modesto:2013jea}. This however is not very appealing since the required value of $m$ depends strongly on the point $x_{\rm in}$ where we set the initial conditions. As we move $x_{\rm in}$ toward $-\infty$, the required graviton mass becomes smaller and smaller, in order to suppress the exponential growth for a longer time. Furthermore, even in this way  it is not possible to obtain a viable DE model. As found in 
\cite{Modesto:2013jea}, with a value $m\simeq 10^{-7}H_0$ one can suppress the growth of the instability during RD, for an evolution starting at a redshift $z\sim 10^6$,  and one can obtain a DE of order of the observed value today, but its EOS parameter today, $w_0$, turns out to be  between $-1.7$ and $-1.5$, which is  not consistent with the present cosmological observations.}
This conclusion extends to all models of the form (\ref{modela1a2}) as long as $b_1\neq 0$, i.e. as long as the operator $\iBox\Gmn$ is present, since its inclusion automatically brings in the function $V(x)$ which is responsible for the instability.
The fact that tensor non-localities generically brings instabilities has also been recently  found, in a different non-local model,  in
\cite{Ferreira:2013tqn}.

\bibliographystyle{utphys}
\bibliography{myrefs_massive}

\end{document}